\documentclass[sn-mathphys-num]{sn-jnl}

\usepackage{graphicx}%
\usepackage{multirow}%
\usepackage{amsmath,amssymb,amsfonts}%
\usepackage{amsthm}%
\usepackage{mathrsfs}%
\usepackage[title]{appendix}%
\usepackage{xcolor}%
\usepackage{textcomp}%
\usepackage{manyfoot}%
\usepackage{booktabs}%
\usepackage{algorithm}%
\usepackage{algorithmicx}%
\usepackage{algpseudocode}%
\usepackage{listings}%
\usepackage{afterpage}%
\usepackage{caption}
\usepackage{tabularx}
\usepackage[utf8]{inputenc}
\usepackage{hyperref}
\usepackage{natbib}
\usepackage{float}

\theoremstyle{thmstyleone}%
\theoremstyle{thmstyletwo}%

\theoremstyle{thmstylethree}%

\begin{document}

\title[Article Title]{Optimizing OpenFaaS on Kubernetes: Comparative Analysis of Language Runtimes and Cluster Distributions}

	\author*[1]{\fnm{Ehsan} \sur{Ataie}}\email{ataie@umz.ac.ir}
	
	\author[1]{\fnm{Mohammadreza} \sur{Pooshani}}\email{m.pooshani17@umail.umz.ac.ir}
	\equalcont{These authors contributed equally to this work.}
	
	\author[2]{\fnm{Hossein} \sur{Aqasizade}}\email{h.aqasizade@gamil.com}
	\equalcont{These authors contributed equally to this work.}
	
	\affil[1]{\orgdiv{Department of Computer Engineering}, \orgname{University of Mazandaran}, \orgaddress{\state{Babolsar}, \country{Iran}}}
	\affil[2]{\orgdiv{Department of Information Engineering and Computer Science}, \orgname{ University of Trento}, \orgaddress{\state{Trento}, \country{Italy}}}


\abstract{Serverless computing, particularly Function-as-a-Service (FaaS), has revolutionized cloud computing by abstracting infrastructure management and enabling dynamic resource allocation. This paper examines the performance and compatibility of OpenFaaS, an open-source serverless platform, when deployed on various Kubernetes distributions, including Kubeadm, K3s, MicroK8s, and K0s.  Moreover, leveraging the CloudLab infrastructure, this study examines the impact of Python, Go, and Node.js programming languages on the performance of Kubernetes-enabled OpenFaaS, specifically when these languages are used to develop functions deployed on the platform. The performance is evaluated and analyzed under various levels of concurrent invocations using several usage-level metrics, such as throughput and CPU usage, as well as responsiveness metrics, such as delay. According to our findings, Go consistently outperforms Python and Node.js in terms of throughput and CPU usage, making it the ideal runtime for serverless applications. Among the Kubernetes distributions, K3s and Kubeadm exhibit superior performance, with Kubeadm maintaining low latency and efficient CPU usage, and K3s demonstrating high throughput. This study provides valuable insights into optimizing the Kubernetes-enabled OpenFaaS platform, highlighting the strengths and trade-offs of different Kubernetes distributions and language runtimes.}




\maketitle

\section{Introduction}\label{sec1}

Cloud computing has transformed a significant portion of the IT industry by making software as a service more appealing and changing how hardware is designed and purchased.\cite{armbrust2009above, aqasizade2024kubernetes, aqasizade2024experimental} Cloud computing is evolving into a more powerful paradigm, with serverless computing emerging as a powerful way to deploy applications and services. Serverless computing is usually based on Function-as-a-Service (FaaS), a model that does not require underlying virtualization to be managed by application developers.\cite{risco2021serverless, shojaee2024memory}
Additionally, it allows developers to provision virtual infrastructure and allocate resources efficiently.\cite{scheuner2020function, yussupov2021faasten} Using this approach provides unparalleled scalability because resources are dynamically allocated according to workload demands, resulting in cost savings and higher resource utilization.\cite{bouizem2023integrating} 

Conducting a comprehensive assessment of serverless computing systems is fundamental to sustaining service reliability, optimizing operational indicators, and controlling execution costs within acceptable thresholds \citep{ghorbian2024survey,kumari2024acpm}. Moreover, such evaluations enable service providers to refine workload scheduling and resource utilization strategies, which can in turn decrease overall infrastructure expenditure and energy consumption. This type of analysis is particularly critical because serverless environments are often intertwined with multiple underlying subsystems, including container orchestration frameworks responsible for managing function instances, and programming language runtimes that execute the deployed functions across the platform.

Various serverless platforms are available, ranging from industry leaders' products, such as AWS Lambda\cite{lambda}, Microsoft Azure Functions(AZF)\cite{azure}, Google Cloud Functions (GCF)\cite{GoogleCloudFunctions}, and IBM Cloud Functions (ICF)\cite{IBMCloudFunctions}, to many open-source alternatives, including OpenFaaS\cite{OpenFaaS}, OpenWhisk\cite{openwhisk}, Kubeless\cite{Kubeless}, Nuclio\cite{nuclio}, Fission\cite{fission}, and Knative\cite{knative}. 
OpenFaaS is renowned  for its significant features, including active community support, high number of active developers, and  potential for extensions. It can be run and managed in a cluster of computers using Kubernetes (K8s), a noted open-source container orchestration platform. K8s offers several advantages for deploying this serverless platform, including resource efficiency due to optimized resource allocation, scalability with elastic capabilities, flexibility in infrastructure selection, stability in deployment and management processes, integration with native cloud tools, and robust community support\cite{liu2024balanced}. 

A K8s distribution refers to a packaged version of the K8s platform, often bundled with additional tools, configurations, and services to simplify deployment, management, and scaling of clusters.
In this study, Kubeadm\cite{Kubeadm}, K3s\cite{k3s}, MicroK8s\cite{microk8s}, and K0s\cite{k0s} are used because of their importance and  popularity in recent research.\cite{chahoud2023demand,basakfog, koziolek2023lightweight,heckmann2023evaluating, kim2023device} Table \ref{tab1} compares these K8s distributions based on their primary features for various deployment environments.

\begin{figure}[t]
	\centering
	\includegraphics[width=.8\textwidth]{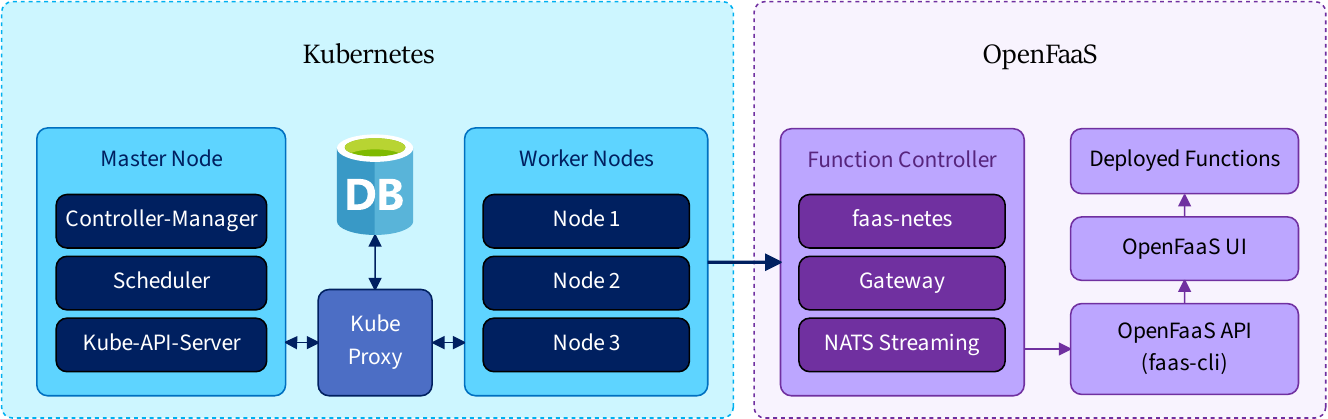}
	\captionsetup{justification=centering}
	\caption{Interdependence between OpenFaaS and Kubernetes}\label{fig1}
\end{figure}

OpenFaaS accommodates a wide range of programming language runtimes, including Python, Go, Java, Ruby, Node.js, and C\#. Among the programming languages supported by OpenFaaS, in this research, we consider Python, Go and Node.js, and assess their performance when they are used to implement functions deployed on the OpenFaaS platform. We use CloudLab \citep{duplyakin2019design}, which is a cloud computing testbed for research and education and has been used as an infrastructure in several related studies \citep{ramanathan2024enabling,ganguly2023multi,qi2023middlenet}. A computationally intensive function is developed and invoked at various concurrency levels using workload generation tools such as Hey \citep{hey} and K6 \citep{k6}. The goal is to analyze the pros and cons of each configuration by accurately evaluating performance measures. Afterward, we analyze and evaluate K8s distributions and programming language runtimes based on various metrics, including requests per second (RPS), average delay, and CPU usage. Briefly, the key contributions of this paper are listed below:

\begin{itemize}\item A comprehensive analysis of OpenFaaS performance across four Kubernetes distributions(Kubeadm, K3s, MicroK8s, and K0s), highlighting their respective strengths and trade-offs.\end{itemize}

\begin{itemize}\item A thorough evaluation of how three popular programming language runtimes(Python, Go, and Node.js] affect critical performance metrics such as throughput, delay, and CPU usage in serverless environments.\end{itemize}

\begin{itemize}\item The use of CloudLab as a controlled, reproducible testbed combined with real-world workload simulation tools (Hey and K6) to ensure the robustness of our evaluation.\end{itemize}

\begin{itemize}\item Identification of Go as the runtime delivering the highest throughput and most efficient CPU usage, alongside insights that reveal K3s and Kubeadm as leading Kubernetes distributions with K3s excelling in throughput and Kubeadm achieving low latency and efficient resource utilization.\end{itemize}

\begin{itemize}\item Practical recommendations for developers and researchers, guiding the selection of the optimal Kubernetes distribution and programming language runtime based on specific deployment requirements.\end{itemize}

The rest of the article is structured as follows: Section \ref{sec2} provides an overview of the related work. Section  \ref{sec3} describes the architecture of the system under study and the software and hardware configurations. Section \ref{sec4} provides the experimental results, including comparisons between different types of K8s distributions and programming language runtimes. Towards the end of the study, Section \ref{sec5} concludes our findings and outlines potential directions for future research.

\begin{table}[htbp]
	\centering
	\caption{Comparative analysis of Kubernetes distributions}
	\label{tab1}
	\footnotesize
	\renewcommand{\arraystretch}{1.2}
	\setlength{\tabcolsep}{4pt}
	
	\begin{tabularx}{\textwidth}{
			>{\raggedright\arraybackslash}p{2.5cm}
			*{4}{>{\raggedright\arraybackslash}X}
		}
		\toprule
		\textbf{Feature} &
		\textbf{Kubeadm} &
		\textbf{K3s} &
		\textbf{K0s} &
		\textbf{MicroK8s} \\
		\midrule
		
		Maintainer
		& K8s SIG
		& Rancher Labs
		& Mirantis
		& Canonical \\
		
		Resource Usage
		& Moderate
		& Low
		& Moderate
		& Low \\
		
		Installation Ease
		& Moderate
		& Easy
		& Easy
		& Easy \\
		
		OS Support
		& Linux
		& Linux, Windows
		& Linux
		& Linux, Windows, macOS \\
		
		Primary Use Case
		& General Purpose
		& Edge, IoT
		& General Purpose
		& Developers, IoT \\
		
		Architecture
		& Full-scale
		& Lightweight
		& Lightweight
		& Lightweight \\
		
		High Availability
		& Supported
		& Supported
		& Supported
		& Supported \\
		
		Snapshots \& Backups
		& External tools
		& Built-in
		& External tools
		& Built-in \\
		
		Community Support
		& Strong
		& Growing
		& Growing
		& Strong \\
		
		Container Network Interface
		& Calico, Flannel, Weave, Cilium
		& Flannel
		& Calico
		& Calico, Flannel \\
		
		Container Runtime
		& Docker, containerd, CRI-O, rkt
		& containerd, Docker
		& containerd, Docker, CRI-O
		& containerd, Docker \\
		
		\bottomrule
	\end{tabularx}
\end{table}

\section{Related Work}\label{sec2}

A review of the literature reveals that serverless computing is a topic of considerable interest to many researchers. With a focus on research relevant to this article, this section aims to provide an overview of studies on serverless computing and K8s. 

Aqasizade et al. \cite{aqasizade2024kubernetes} evaluated the performance of four Kubernetes distributions including Kubeadm, K3s, MicroK8s, and K0s. Their experiments examined two virtualization modes, HVM and PV, supported by the Xen hypervisor. Additionally, they compared two container runtimes, Docker and Containerd, assessing their performance across disk-intensive and CPU-intensive workloads. After identifying the optimal Xen mode and container runtime, they evaluated the Kubernetes distributions using various performance metrics, including request rate, CPU utilization, and scaling behavior.

According to Koziolek et al. \cite{koziolek2023lightweight}, MicroK8s, K3s, K0s, and MicroShift are lightweight K8s distributions, and they discovered that K3s and K0s achieved marginally higher control plane throughputs in stress scenarios, whereas MicroShift achieved higher data plane throughputs.
Kjorveziroski et al. \cite{kjorveziroski2022kubernetes} have examined the performance of three K8s distributions: Kubespray, K3s, and MicroK8s using serverless benchmarks using OpenFaaS. Especially in terms of reducing deployment time and complexity, lightweight distributions like K3s and MicroK8s outperformed Kubespray. Even though there were some performance variations in extreme load tests, MicroK8s and K3s generally performed similarly. 

A study conducted by Mohanty et al. \cite{mohanty2018evaluation} examined open-source serverless platforms such as Kubeless and OpenFaaS, as well as Fission and OpenWhisk. Based on average delay and success rates under different load conditions, the authors evaluated the features and architectures of each platform. They found that Kubeless provided a more stable performance than Fission and OpenFaaS. In an edge computing environment for Internet of Things (IoT) applications, Palade et al. \cite{8817155} have evaluated four open-source serverless platforms, namely Kubeless, OpenWhisk, OpenFaaS, and Knative. Using JMeter, the authors assessed average delay, operational capacity, and success rates of deployed functions under various workloads and found that Kubeless demonstrated superior performance in comparison with other platforms.

Balla et al. \cite{9163456} have evaluated how Python, Node.js, and functions on OpenFaaS, Kubeless, Fission, and Knative perform in terms of execution times. Researchers found a performance difference between compute-intensive and I/O-intensive workloads when using OpenFaaS due to session timeouts, which directed all requests to a single function instance, which negatively affected scale performance. Moreover, when there are many concurrent users, Python execution in Kubeless is unresponsive to health check messages, but this can be fixed by adjusting the runtime parameters. Li et al. \cite{Li_2021} have examine key characteristics of open-source serverless platforms like Nuclio, OpenFaaS, Knative, and Kubeless and address performance gaps in service exporting and auto-scaling. OpenFaaS and Knative experienced delays due to multiple components queuing at the same time, while Nuclio had the lowest latency at 99 percentile. OpenFaaS and Nuclio, two open-source serverless platforms, have been evaluated by Decker et al. \cite{a15070234} in a cluster of high-performance computing (HPC). According to the researchers, Nuclio had 1.5x the data throughput of OpenFaaS, and existing platforms need improvement for optimal HPC usage. Enhancing performance was made possible by integrating user functions with function instances.

Lee et al. \cite{8457830} have evaluated serverless computing's performance in processing distributed data through parallel function invocation. Study results demonstrated the effectiveness of serverless in specific data center applications, such as IoT-driven live migrations. Wen et al. \cite{https://doi.org/10.1002/smr.2394} have described four enterprise serverless computing platforms Comprising AWS Lambda, Google Cloud Functions, Azure Functions, and Alibaba Cloud Function Compute. It has been found that too many function instances can cause poor performance in a single virtual machine. Yu et al. \cite{yu2020characterizing} have introduced an open-source benchmark suite named ServerlessBench to assess the efficiency of communication, launch latency, stateless overhead, and performance isolation of AWS Lambda, OpenWhisk, and Fn platforms. The authors then examined the impact of the introduced benchmark on AWS Lambda, OpenWhisk, and Fn platforms. According to Wang et al. \cite{wang2018peeking}, three platforms, AWS Lambda, Google Cloud Function, and Microsoft Azure, were compared for cold start latency by measuring the time difference between function invocation and function initiation. Also, the authors investigated how programming language runtime, memory, and the number of function instances affect cold start latency. It was found that memory and programming language runtime had a significant effect on cold start latency. Among the languages tested on the AWS Lambda platform, Python has the lowest average cold start latency, while Java functions have higher latency. 

Figiela et al. \cite{figiela2018performance} have conducted an evaluation of the AWS Lambda, Microsoft Azure Functions, Google Cloud Function, and IBM Cloud Function platforms. They introduced a new benchmarking framework that uses both the Serverless framework and HyperFlow. Among cloud service providers, significantly different execution patterns were found. AWS Lambda execution was related to memory but slower at times. Google Cloud Function execution performed better with more memory. Javed et al. \cite{javed2022serverless} evaluated multiple serverless platforms, including OpenFaaS, AWS Greengrass, and Apache OpenWhisk, when installed on edge 

Ataie et al. \cite{ataie2024} investigated the performance of three open-source serverless platforms, including OpenFaaS, Nuclio, and Fission in terms of programming languages. They focused on Python and Node.js functions, evaluating the platforms based on response time, throughput, and total data exchanged. The results indicated that Python was the better choice for OpenFaaS, while Node.js performed better on Fission, especially under high concurrency. In contrast, Nuclio outperformed both platforms regardless of the programming language used. 

Goethals et al. \cite{goethals2019fledge} introduced FLEDGE, a Kubernetes-compatible edge container orchestrator designed for low-resource environments. They examined several key factors for efficient container orchestration, including the choice of container runtime and the implementation of container networking. Additionally, they conducted multiple evaluations, comparing FLEDGE with K3s and Kubernetes. Böhm et al. \cite{bohm2021profiling} evaluated lightweight Kubernetes distributions, MicroK8s and K3s, which aim to simplify K8s deployment in resource-constrained environments. Their experiment compared resource usage and time consumption during key lifecycle operations like starting, stopping, and adding nodes. The results showed that K3s had similar resource consumption to K8s but performed better when starting and adding nodes. MicroK8s, however, showed higher resource usage and longer operation times. Koukis et al. \cite{koukis2024performance} evaluated the networking performance of four Kubernetes implementations including vanilla K8s, K3s, K0s, and MicroK8s and focusing on five Container Network Interface (CNI) plugins such as Flannel, Calico, Cilium, Kube-router, and Kube-ovn. Their assessment revealed that lightweight K8s distributions do not always lead to resource savings in terms of CPU, memory, or throughput, even with different CNI plugins.

Tzenetopoulos et al. \cite{tzenetopoulos2021faas} proposed a methodology for evaluating serverless frameworks deployed on hybrid edge-cloud clusters. Their approach focused on key performance factors of serverless computing and systematically evaluated three open-source frameworks including OpenFaaS, OpenWhisk, and Lean OpenWhisk. Ustiugov et al. \cite{ustiugov2021analyzing} introduced STeLLAR, a benchmarking framework for serverless performance analysis. They also evaluated AWS Lambda, Google Cloud Functions, and Azure Functions, identifying storage accesses and bursty traffic as primary contributors to latency variability. Interestingly, the choice of language runtime had little impact on latency. Trieu et al. \cite{trieu2022performance} conducted a comprehensive performance analysis of a serverless edge computing system using popular open-source platforms such as Kubeless, OpenFaaS, Fission, and funcX. They evaluated various programming languages, workloads, and concurrent user counts, particularly focusing on machine learning workloads.

In contrast to earlier research, the present work concentrates on OpenFaaS compatibility and performance in relation with various K8s distributions, including Kubeadm, K3s, MicroK8s, and K0s, by using its robust container orchestration capabilities. Furthermore, this research provides valuable insights into programming language runtime performance, including Go, Python, and Node.js. The main objective is to identify which combinations of K8s distributions and language runtimes deliver the best performance for serverless applications running on OpenFaaS. Table \ref{tab2} compares our work with others based on various criteria.

\begin{table}[htbp]
	\centering
	\caption{Comparison of our work with related work}
	\label{tab2}
	\scriptsize
	\renewcommand{\arraystretch}{1.2}
	\setlength{\tabcolsep}{3pt}
	
	\begin{tabularx}{\textwidth}{
			>{\raggedright\arraybackslash}p{1.65cm}
			>{\raggedright\arraybackslash}p{1.7cm}
			>{\raggedright\arraybackslash}p{1.8cm}
			>{\raggedright\arraybackslash}p{1.45cm}
			>{\raggedright\arraybackslash}X
			>{\raggedright\arraybackslash}p{1.25cm}
		}
		\toprule
		\textbf{Related Work} &
		\textbf{Serverless Platform} &
		\textbf{Distribution} &
		\textbf{PL Runtime} &
		\textbf{Evaluation Metrics} &
		\textbf{Measurement Tool} \\
		\midrule
		
		Aqasizade et al. \cite{aqasizade2024kubernetes}
		& OpenFaaS
		& Kubeadm, K3s, MicroK8s, K0s
		& Python
		& Response time, throughput, auto-scaling, CPU usage
		& K6 \\
		
		Koziolek et al. \cite{koziolek2023lightweight}
		& N/A
		& MicroK8s, K3s, K0s, MicroShift
		& N/A
		& CPU, memory, disk usage
		& Netdata \\
		
		Kjorveziroski et al. \cite{kjorveziroski2022kubernetes}
		& OpenFaaS
		& KubeSpray, MicroK8s, K3s
		& Python
		& Container startup, application execution, response time, throughput
		& Hey \\
		
		Mohanty et al. \cite{mohanty2018evaluation}
		& Kubeless, OpenFaaS, Fission
		& GKE
		& Go
		& Response time, request success rate
		& AB \\
		
		Palade et al. \cite{8817155}
		& Kubeless, OpenWhisk, OpenFaaS, Knative
		& Kubeadm
		& Node.js
		& Response time, throughput, request success rate, deployment
		& JMeter \\
		
		Balla et al. \cite{9163456}
		& OpenFaaS, Kubeless, Fission, Knative
		& Kubeadm
		& Python, Node.js, Go
		& Completion time, auto-scaling
		& Hey \\
		
		Li et al. \cite{Li_2021}
		& Knative, Kubeless, Nuclio, OpenFaaS
		& Kubeadm
		& Python
		& Response time, throughput, auto-scaling, resource usage
		& Wrk \\
		
		Decker et al. \cite{a15070234}
		& OpenFaaS, Nuclio
		& Kubeadm
		& Python, Go, C++
		& Images per second, throughput
		& N/A \\
		
		Lee et al. \cite{8457830}
		& Lambda, AZF, GCF, ICF
		& Kubeadm
		& Python, Node.js, Java, C\#
		& Response time, throughput, runtime overhead, CPU, I/O
		& N/A \\
		
		Wen et al. \cite{https://doi.org/10.1002/smr.2394}
		& Lambda, AZF, GCF, Alibaba
		& Public clouds
		& Python, Node.js, Java
		& Cold start, resource efficiency
		& TBS \\
		
		Yu et al. \cite{yu2020characterizing}
		& Lambda, OpenWhisk, Fn
		& Public clouds
		& Python, Node.js, Ruby, C
		& Response time, stateless overhead
		& N/A \\
		
		Wang et al. \cite{wang2018peeking}
		& Lambda, AZF, GCF
		& Public clouds
		& Python, Node.js, Java
		& Memory, number of function instances, cold start
		& N/A \\
		
		Figiela et al. \cite{figiela2018performance}
		& Lambda, AZF, GCF, ICF
		& Public clouds
		& Elixir, Node.js
		& Memory, number of function instances, cold start
		& N/A \\
		
		Javed et al. \cite{javed2022serverless}
		& Lambda, AZF, AWS Greengrass, OpenWhisk, OpenFaaS
		& Kubeadm, Docker Swarm
		& Node.js
		& Response time, request success rate
		& JMeter \\
		
		Ataie et al. \cite{ataie2024}
		& OpenFaaS, Nuclio, Fission
		& Kubeadm
		& Python, Node.js
		& Response time, throughput, total data
		& Hey, Wrk \\
		
		Goethals et al. \cite{goethals2019fledge}
		& N/A
		& K8s, K3s, FLEDGE
		& N/A
		& Memory, storage usage
		& df, pmap \\
		
		Bohm et al. \cite{bohm2021profiling}
		& N/A
		& K8s, MicroK8s, K3s
		& N/A
		& CPU, memory, disk usage
		& Netdata \\
		
		Koukis et al. \cite{koukis2024performance}
		& N/A
		& Vanilla K8s, K3s, K0s, MicroK8s
		& N/A
		& Throughput, CPU, memory usage
		& N/A \\
		
		Tzenetopoulos et al. \cite{tzenetopoulos2021faas}
		& OpenFaaS, OpenWhisk, Lean OpenWhisk
		& K3s
		& Python, Node.js, Go, Ruby
		& Idle state, cold start, concurrent invocation, payload transfer, auto-scaling
		& N/A \\
		
		Ustiugov et al. \cite{ustiugov2021analyzing}
		& Lambda, AZF, GCF
		& Public clouds
		& Python, Go
		& Response time, cold start, data communication delays
		& STeLLAR \\
		
		Trieu et al. \cite{trieu2022performance}
		& Kubeless, OpenFaaS, Fission, funcX
		& MicroK8s
		& Python, Node.js
		& Response time, auto-scaling, request success rate, training time
		& JMeter \\
		
		\textbf{Our Work}
		& \textbf{OpenFaaS}
		& \textbf{Kubeadm, K3s, MicroK8s, K0}
		& \textbf{Python, Go, Node.js}
		& \textbf{Response time, throughput, auto-scaling, CPU usage}
		& \textbf{Hey, K6} \\
		
		\bottomrule
	\end{tabularx}
\end{table}

\section{Research Methodology}\label{sec3}
This section details the hardware and software configurations, workload generation methods, and K8s distributions used in this study.

\subsection{Hardware}\label{sec3-1}
\begin{figure}[h]
	\centering
	\includegraphics[width=.6\textwidth]{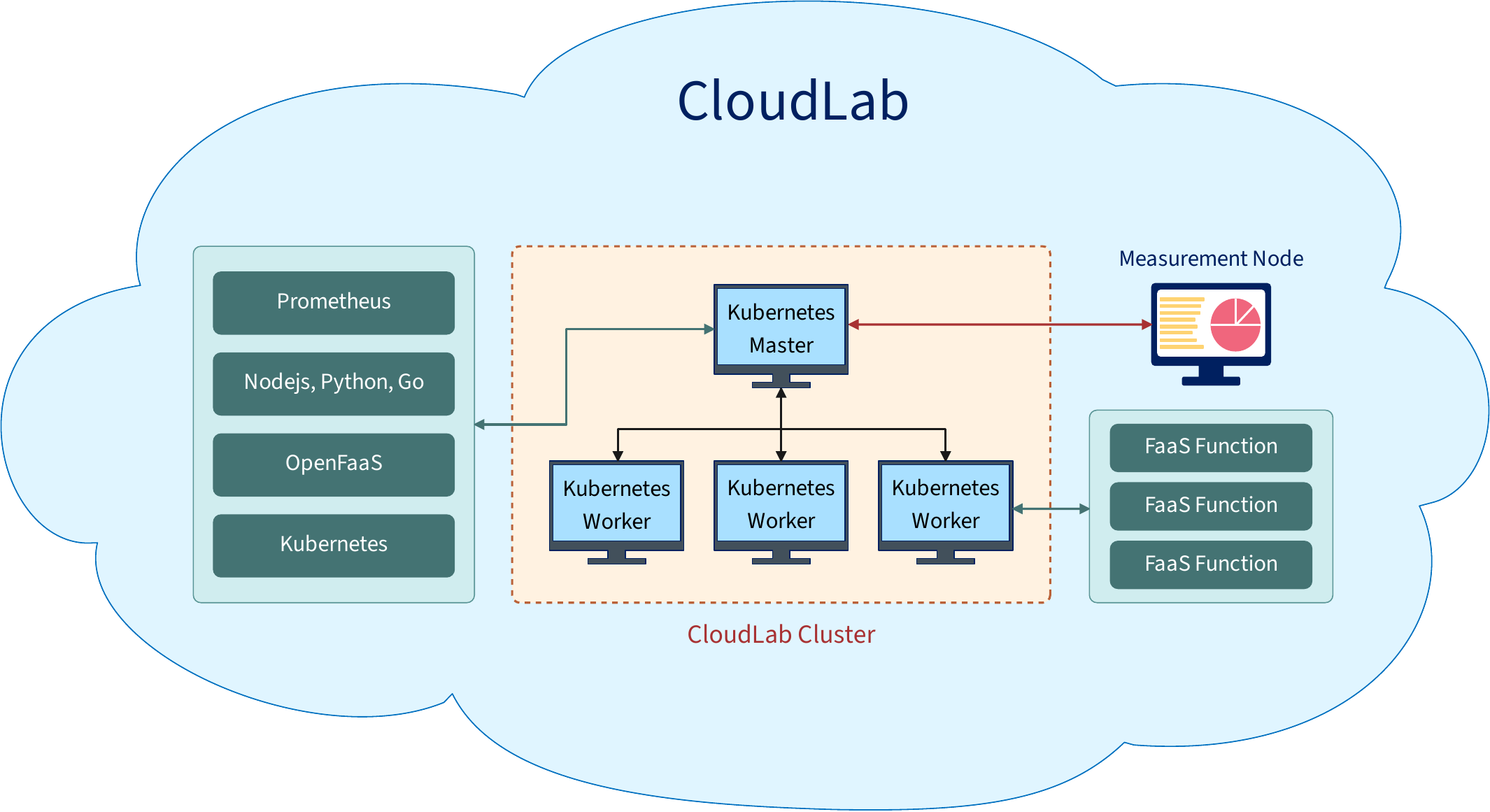}
	\captionsetup{justification=centering}
	\caption{Deployment architecture}\label{fig2}
\end{figure}

The research is conducted on CloudLab, which is renowned for its robust resource allocation and clustering capabilities.
CloudLab gives researchers control over a slice of the facility, allowing them to run tailored cloud software. 
The experiments are carried out on five c6620 machines (four nodes for OpenFaaS and one node for measurement), which are outfitted with dual Intel Xeon E5-2650v2 processors (each featuring 8 cores at 2.6 GHz), 64 GB of RAM, a 1 TB disk, and an Intel X520 PCIe dual-port 10Gb Ethernet NIC.
Figure \ref{fig2} and Table \ref{tab3} show the deployment architecture and hardware details, respectively.

\begin{table}[!h]
	\centering
	\caption{Hardware specifications in CloudLab experiments}\label{tab3}
	\begin{tabular}{lll}
		\toprule
		\textbf{Criterion} & \textbf{Values}  \\
		\midrule
		Number of nodes    & 5 (1 master, 3 workers, 1 measurement)  \\
		Processor & Intel\textsuperscript{\textregistered} Xeon\textsuperscript{\textregistered} E5-2650v2 \\
		Memory    & 64 GB  \\
		Disk    & 1 TB \\
		Network Card    & Intel X520 PCIe dual-port 10Gb  \\
		\bottomrule
	\end{tabular}
\end{table}

\vspace{-0.6cm}

\subsection{Software}\label{sec3-2}
All experiments are conducted on Ubuntu Server 20.04\cite{ubuntu}. Additionally, The OpenFaaS platform version 0.27.1 is installed on all K8s distributions using Helm\cite{helm} version 3.13.3. 
Calico\cite{calico} is the container network interface (CNI) plugin that is installed in all distributions of K8s to ensure fairness. Furthermore, we choose Python, Go, and Node.js as programming languages to be investigated for a variety of reasons, including alignment with existing research and popularity, language diversity and applicability, industry relevance, and community support. Moreover, we visualize the results using Prometheus\cite{prometheus} and Grafana\cite{grafana}, which are installed on the measurement node. A query is written to obtain values specific to the OpenFaaS and OpenFaaS-Fn namespaces which can be accessed on our GitHub \cite{pk8s}. 
Table \ref{tab4} summarizes the software configurations in our experiments.

\begin{table}[h]
	\centering
	\caption{Software specifications of CloudLab experiments}\label{tab4}
	\begin{tabular}{lll}
		\toprule
		\textbf{Criterion} & \textbf{Values}  \\
		\midrule
		Operating System    & Ubuntu Server 20.04 \\
		Kubernetes Version    & 1.27.2  \\
		CNI Plugin    & Calico v3.16  \\
		OpenFaaS Version    & 0.27.1  \\
		Helm Version  & 3.13.3   \\
		Prometheus Version  & 2.47.0   \\
		Grafana Version  & 9.5.1   \\
		Programming Language Runtimes & Node.js, Python, Go \\
		OpenFaaS Function & Factorial 500 \\
		\bottomrule
	\end{tabular}
\end{table}

\subsection{Workload Generation}\label{sec3-3} 
A load test tool called Grafana K6 is used in this study in order to generate workloads on the serverless platform and clustering services. However, we repeat the experiments with Hey tool to ensure accuracy.
These tools generate HTTP requests by calling functions deployed on the serverless platform.
The function implemented in Python, Go and Node.js is a \textit{factorial} function with parameter 500 which pressures the processor.
Requests are sent at various concurrency levels, i.e., 5, 10, and 20, for three minutes.
It turns out that after generating this load, it takes nearly two minutes for the input queue to become empty. Consequently, we introduce a 5-minute delay between subsequent runs to prevent experiments from interfering with one another. To ensure maximum accuracy, all measurements are repeated independently and separately 10 times.

\subsection{Kubernetes distributions}\label{sec3-4}
Four K8s distributions are investigated and compared in this paper, namely Kubeadm, K3s, MicroK8s, and K0s, using a variety of metrics. Kubeadm is a command-line tool used to set up K8s clusters on Linux. It simplifies booting up new clusters. A lightweight distribution like K3s, on the other hand, only contains the required components with the goal of reducing CPU and memory usage. This distribution is primarily designed to operate in constrained environments, such as edge computers and IoT devices. The MicroK8s distribution from Canonical is lightweight and based on K3s. For development and testing, it provides a straightforward K8s distribution. K0s, the latest K8s distribution from Mirantis, simplifies and reduces orchestration complexity by utilizing  K8s' standard design across both local and cloud networks.
To maintain fairness, all K8s distributions are standardized to version 1.27.2.

\section{Evaluation Results}\label{sec4}

This section presents the results of performance evaluation. 
In Section \ref{subsec4-1}, three programming language runtimes with four K8s distributions are assessed in terms of throughput and average delay as performance measures. Next, in Section \ref{subsec4-2}, the CPU usage of these programming language runtimes is analyzed. 
Afterwards, Section \ref{subsec4-3} compares K8s distributions in terms of different metrics when Go is chosen as the programming language, and Section \ref{subsec4-4} provides a discussion of the results and their analysis.

\subsection{RPS and Average Delay}\label{subsec4-1}

\begin{figure}[!h]
	\centering
	\begin{minipage}{0.48\textwidth}
		\centering
		\includegraphics[width=\textwidth]{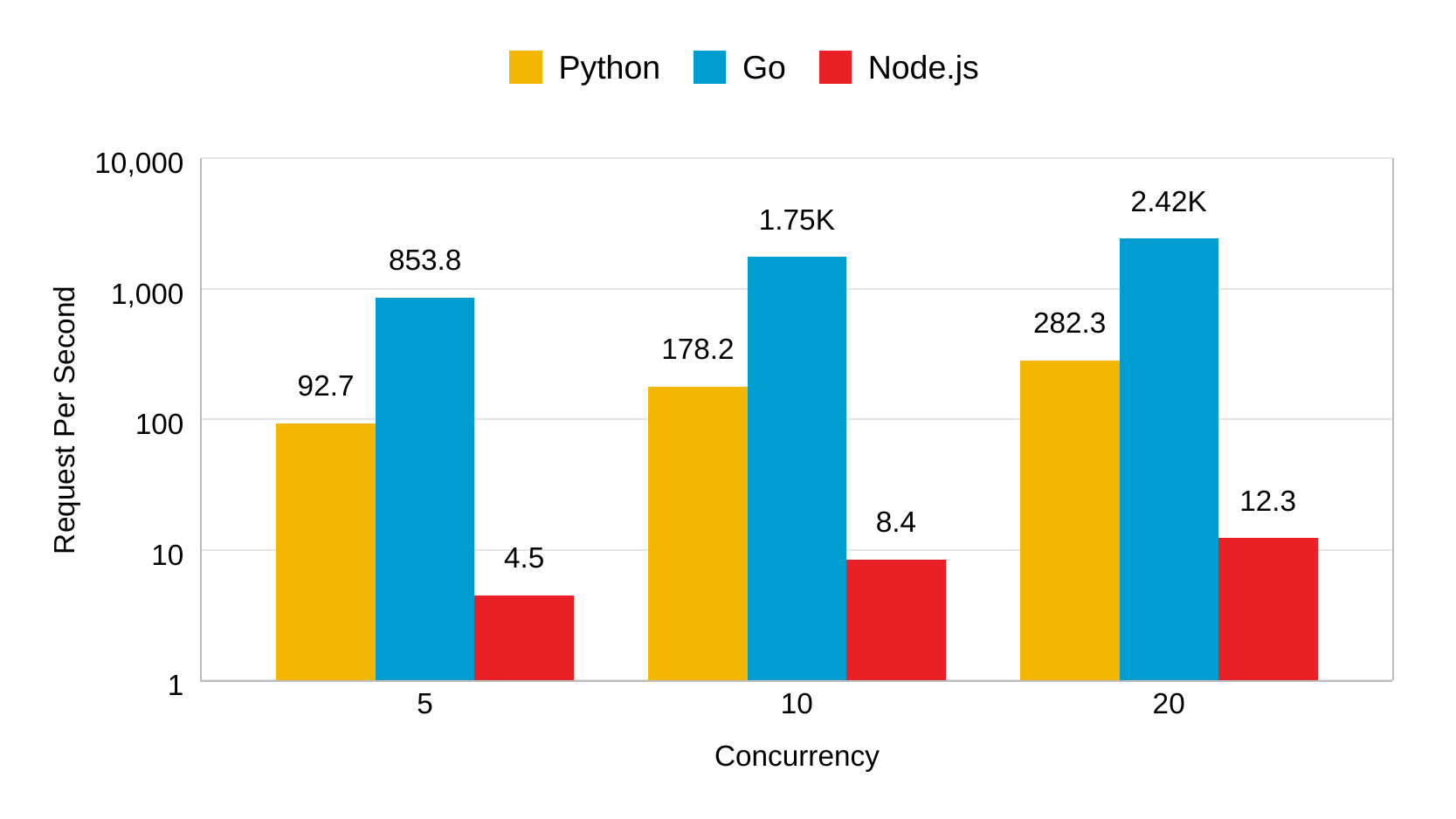}
		\caption{Requests per second for Kubeadm}
		\label{fig3}
	\end{minipage}%
	\hfill
	\begin{minipage}{0.48\textwidth}
		\centering
		\includegraphics[width=\textwidth]{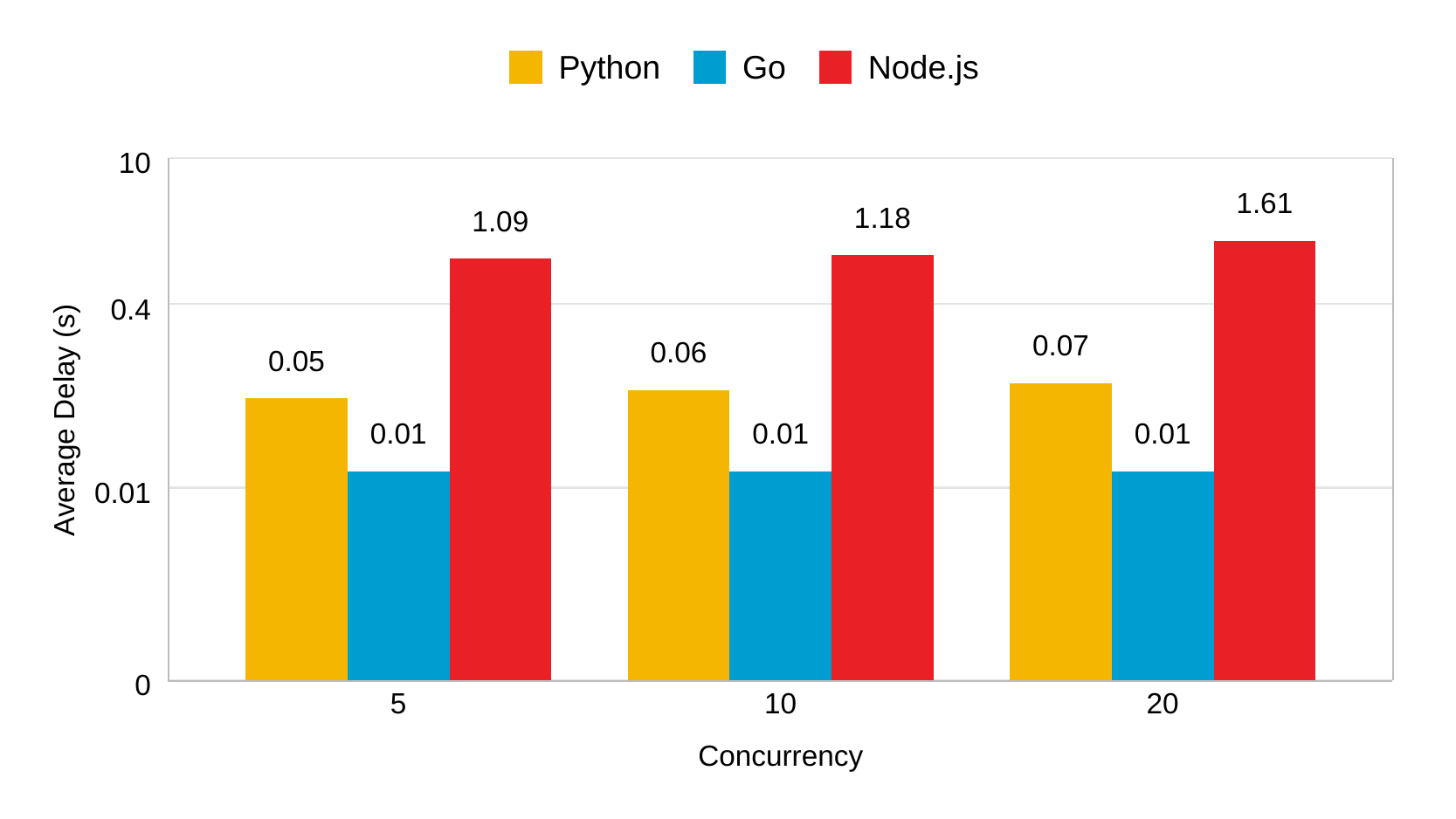}
		\caption{Average delay for Kubeadm}
		\label{fig4}
	\end{minipage}
\end{figure}

\begin{figure}[!h]
	\centering
	\begin{minipage}{0.48\textwidth}
		\centering
		\includegraphics[width=\textwidth]{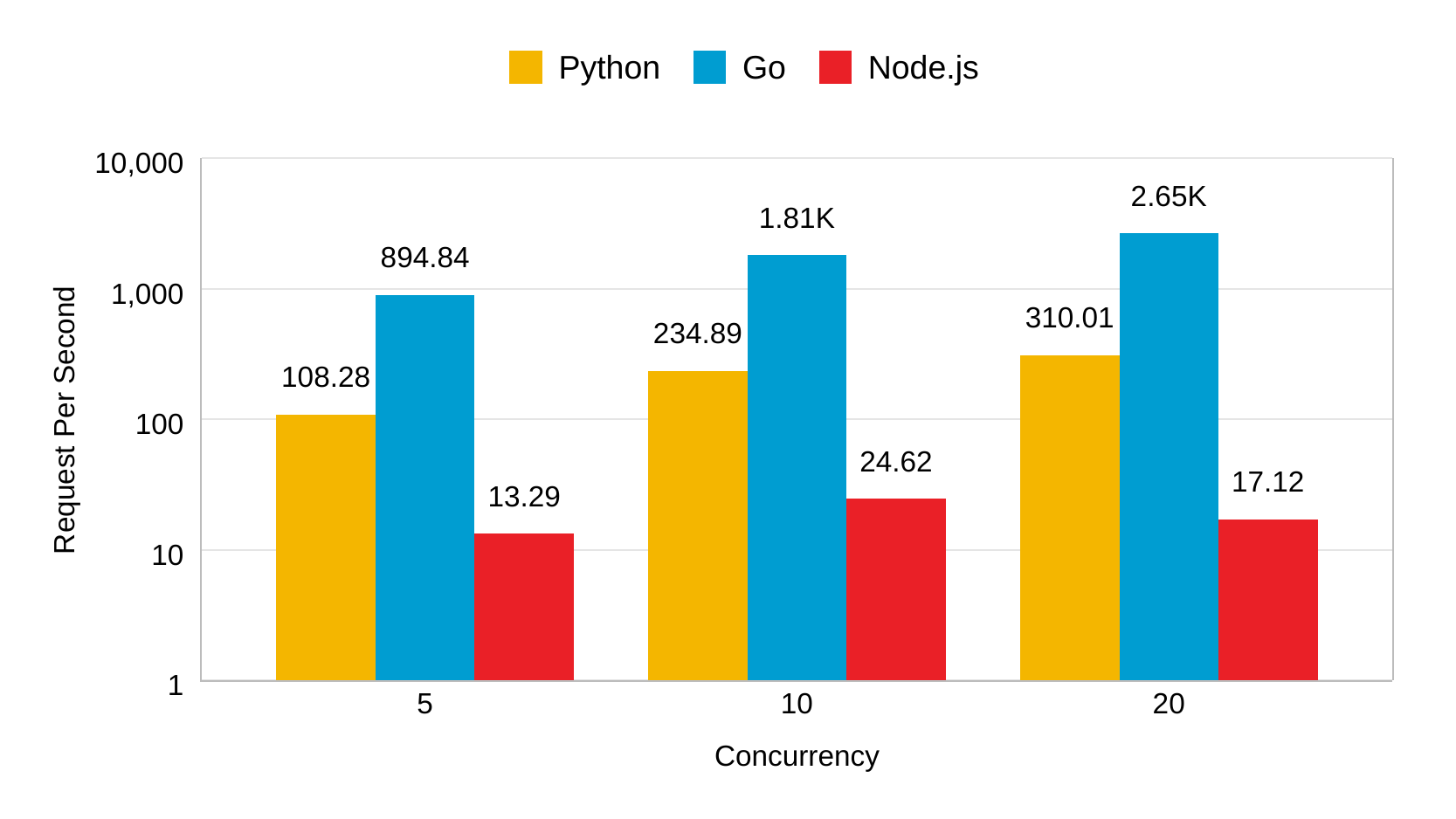}
		\caption{Requests per second for K3s}
		\label{fig5}
	\end{minipage}%
	\hfill
	\begin{minipage}{0.48\textwidth}
		\centering
		\includegraphics[width=\textwidth]{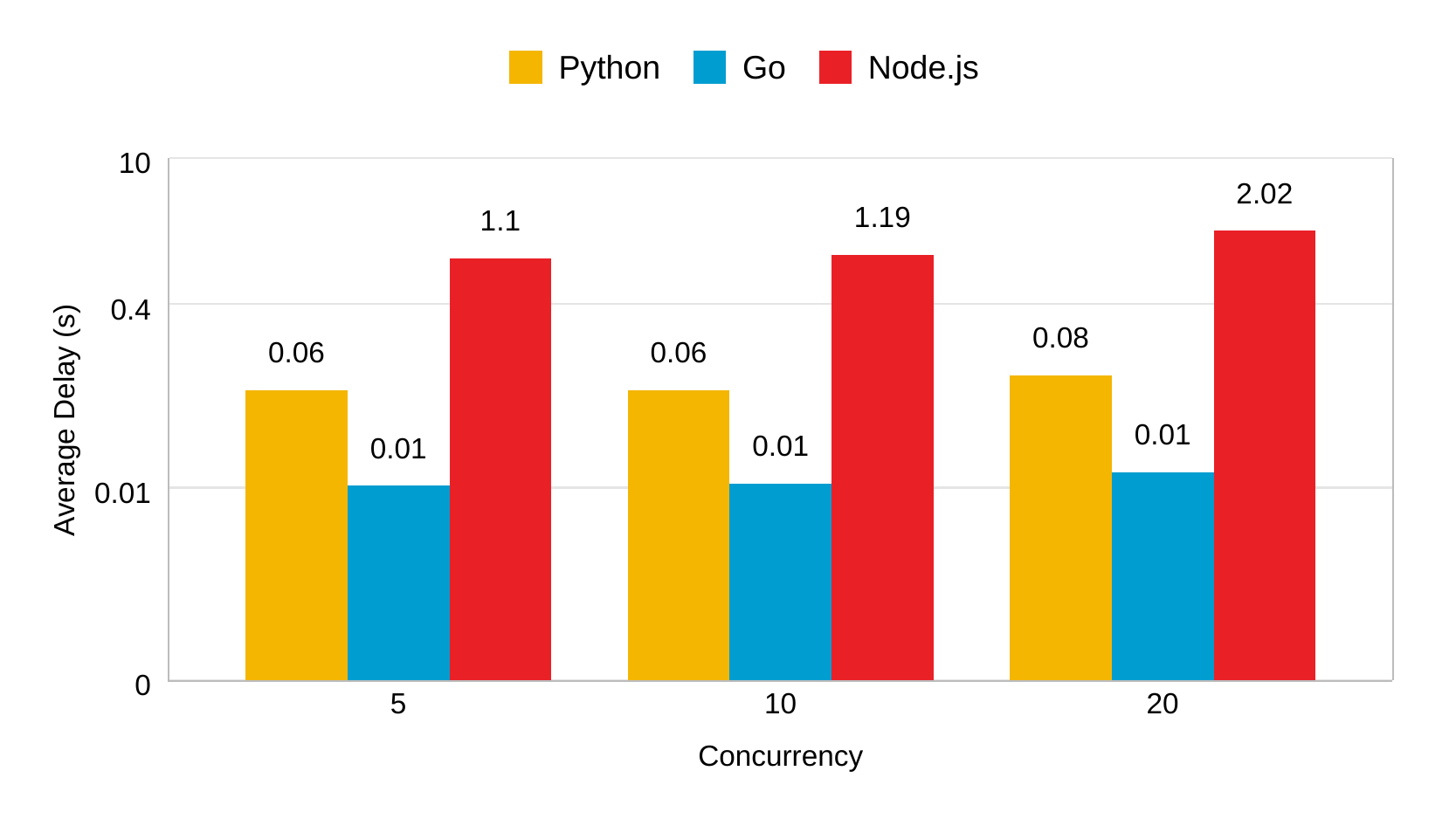}
		\caption{Average delay for K3s}
		\label{fig6}
	\end{minipage}
\end{figure}

\begin{figure}[!h]
	\centering
	\begin{minipage}{0.48\textwidth}
		\centering
		\includegraphics[width=\textwidth]{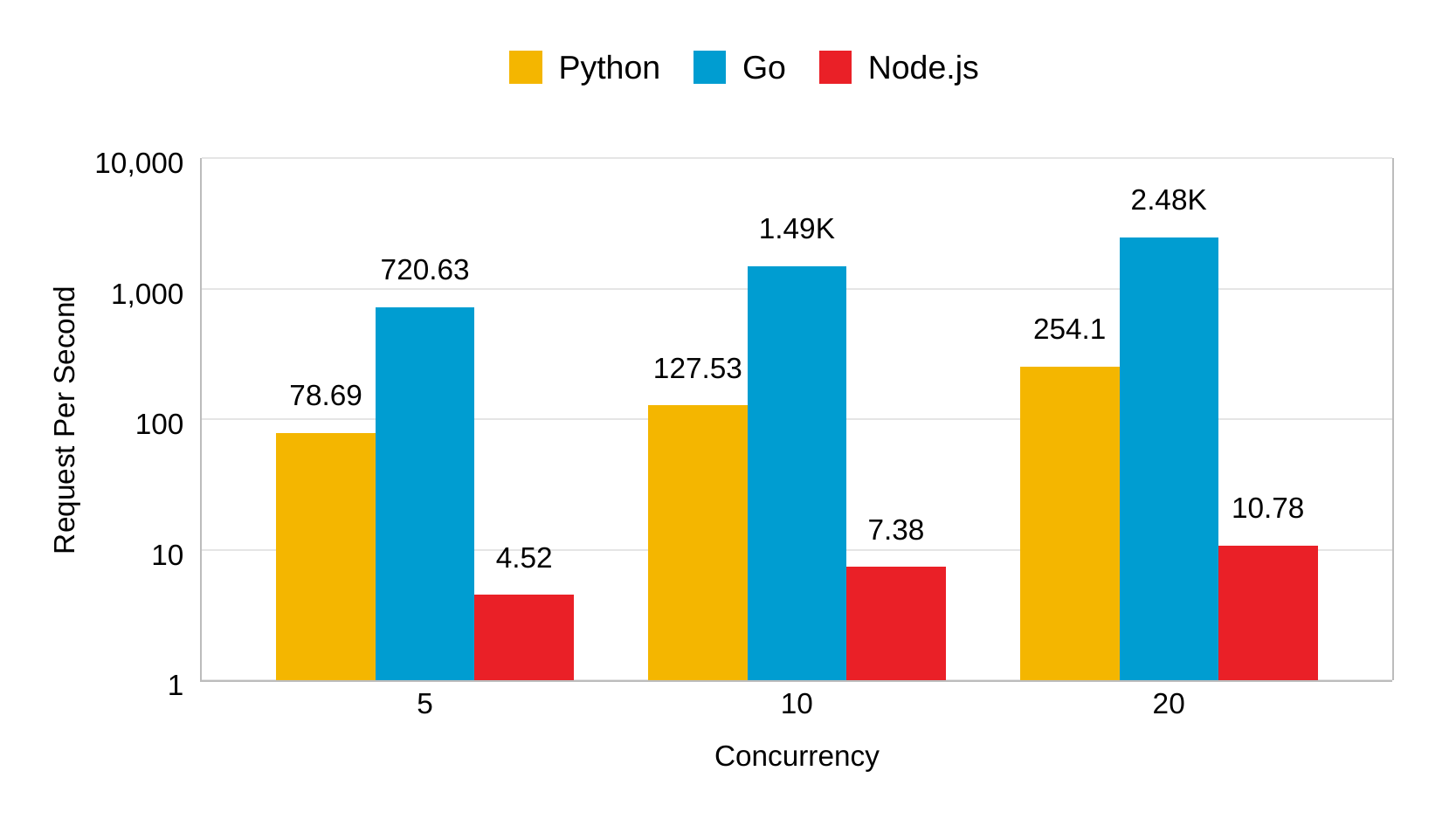}
		\caption{Requests per second for MicroK8s}
		\label{fig7}
	\end{minipage}%
	\hfill
	\begin{minipage}{0.48\textwidth}
		\centering
		\includegraphics[width=\textwidth]{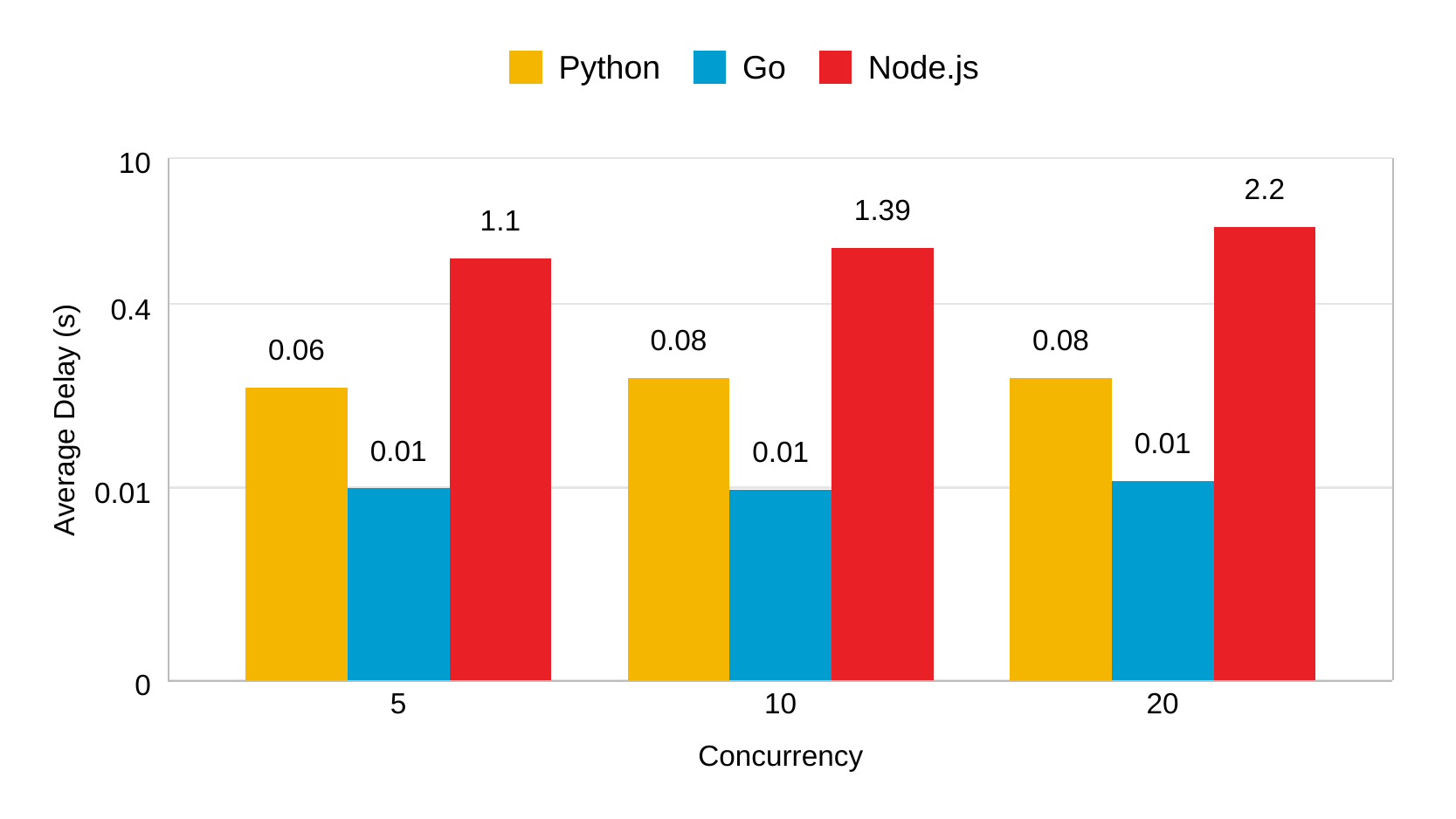}
		\caption{Average delay for MicroK8s}
		\label{fig8}
	\end{minipage}
\end{figure}

\begin{figure}[!h]
	\centering
	\begin{minipage}{0.48\textwidth}
		\centering
		\includegraphics[width=\textwidth]{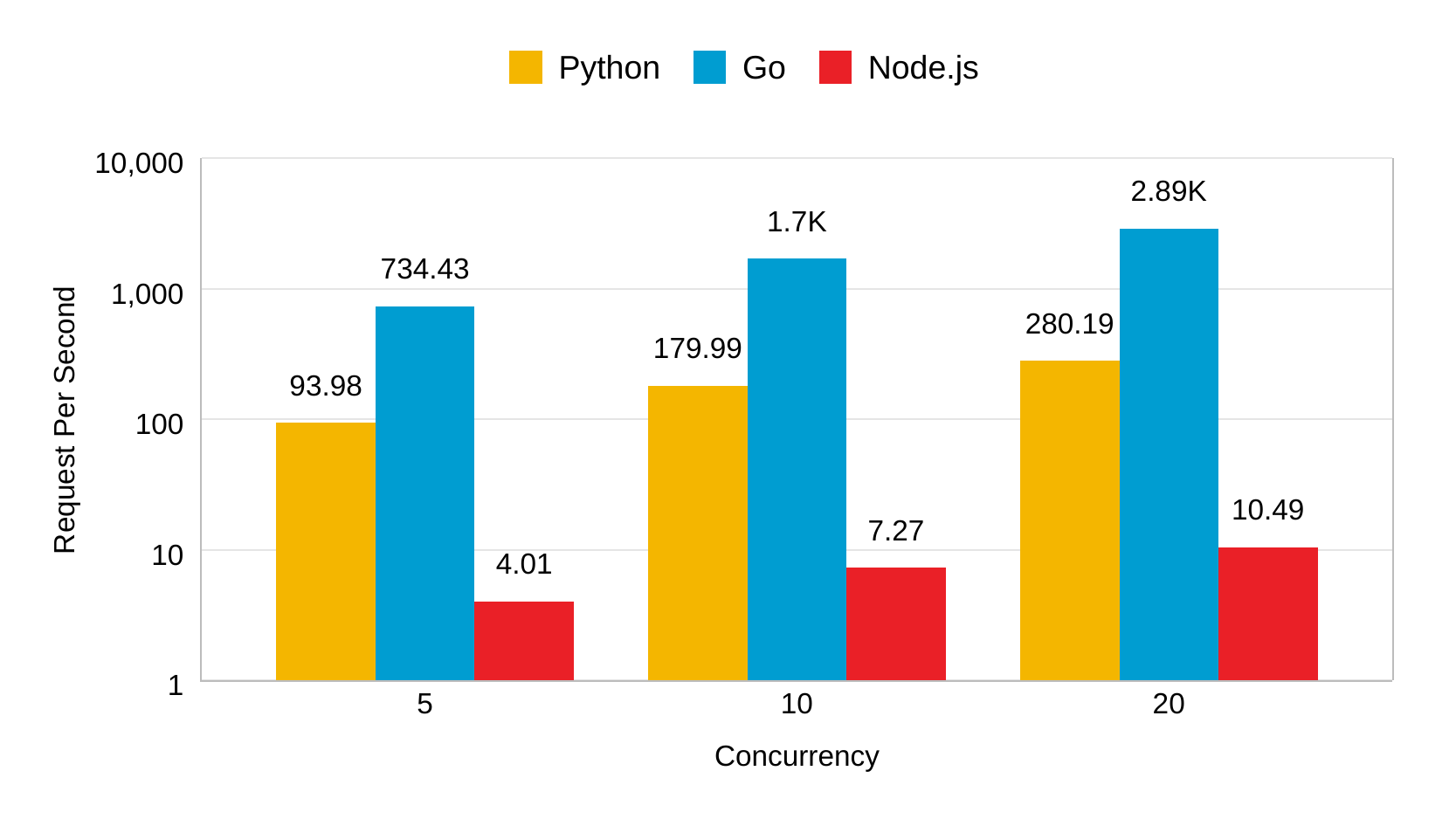}
		\caption{Requests per second for K0s}
		\label{fig9}
	\end{minipage}%
	\hfill
	\begin{minipage}{0.48\textwidth}
		\centering
		\includegraphics[width=\textwidth]{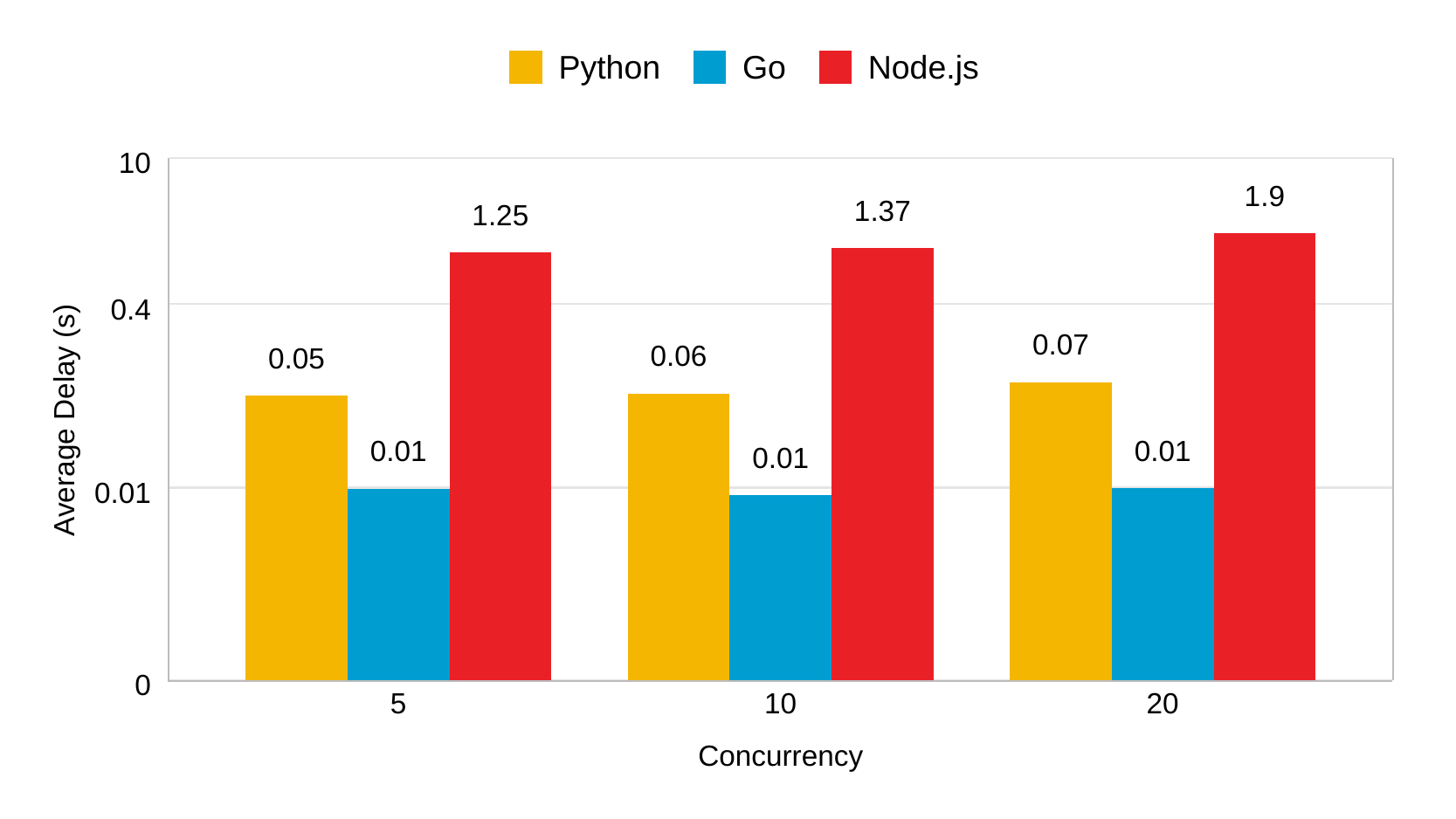}
		\caption{Average delay for K0s}
		\label{fig10}
	\end{minipage}
\end{figure}

Figures \ref{fig3} through \ref{fig10} depict the RPS and average delay for various programming language runtimes on Kubeadm, K3s, MicroK8s, and K0s, tested with 5, 10, and 20 concurrent users, all presented on a logarithmic scale.
The analysis of K8s distributions performance using Python, Go, and Node.js as programming language reveals several interesting trends:

In terms of RPS, Go language performs better than the other two languages regardless of K8s distribution and concurrency level. Python is second, while Node.js is third.
Regardless of one exception, as concurrency increases, throughput improves for all languages and K8s distributions.
It seems that Go is the most sensitive language to concurrency levels, especially when combined with K0s distribution. On the other hand, Node.js is the least sensitive when paired with K3s.
Although K8s distributions perform differently depending on the language and the concurrency level, K3s ranks on average as the winner, Kubeadm comes in second, K0s comes in third, and MicroK8s comes in fourth. 

Based on the delay, Go once again outperforms other languages regardless of K8s distribution and concurrency level. Python is ranked second, and Node.js is ranked third.
While for Node.js and Python the average delay increases when the concurrency level increases, for Go, it remains almost constant across all K8s distributions.
Kubeadm absolutely provides a shorter delay in contrast to other distributions. K3s and K0s are ranked next acting almost the same, and MicroK8s is ranked fourth on average.

\subsection{CPU Usage}\label{subsec4-2}

\begin{figure}[!h]
	\centering
	\begin{minipage}{0.48\textwidth}
		\centering
		\includegraphics[width=\textwidth]{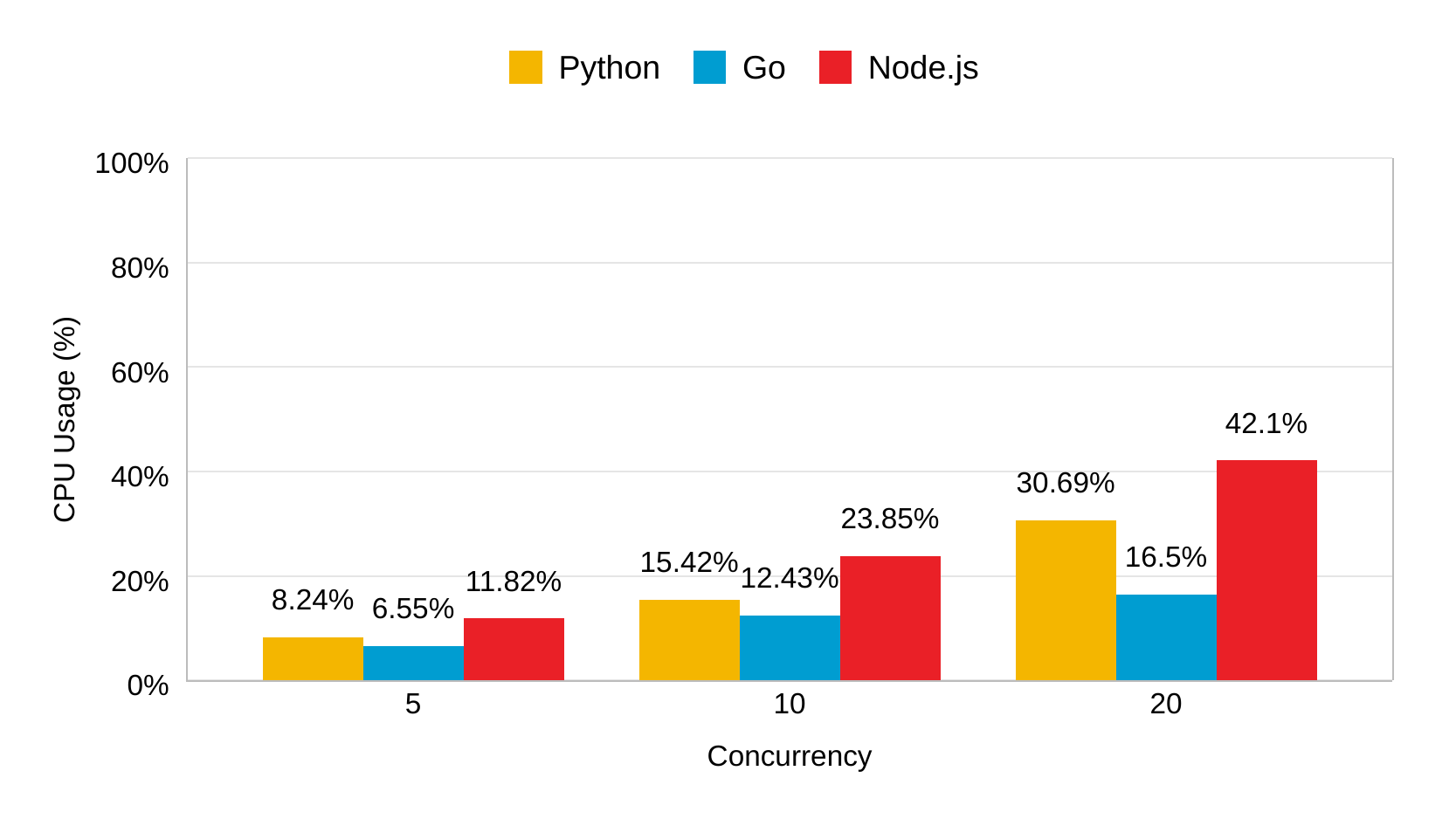}
		\caption{Kubeadm CPU usage}
		\label{fig11}
	\end{minipage}%
	\hfill
	\begin{minipage}{0.48\textwidth}
		\centering
		\includegraphics[width=\textwidth]{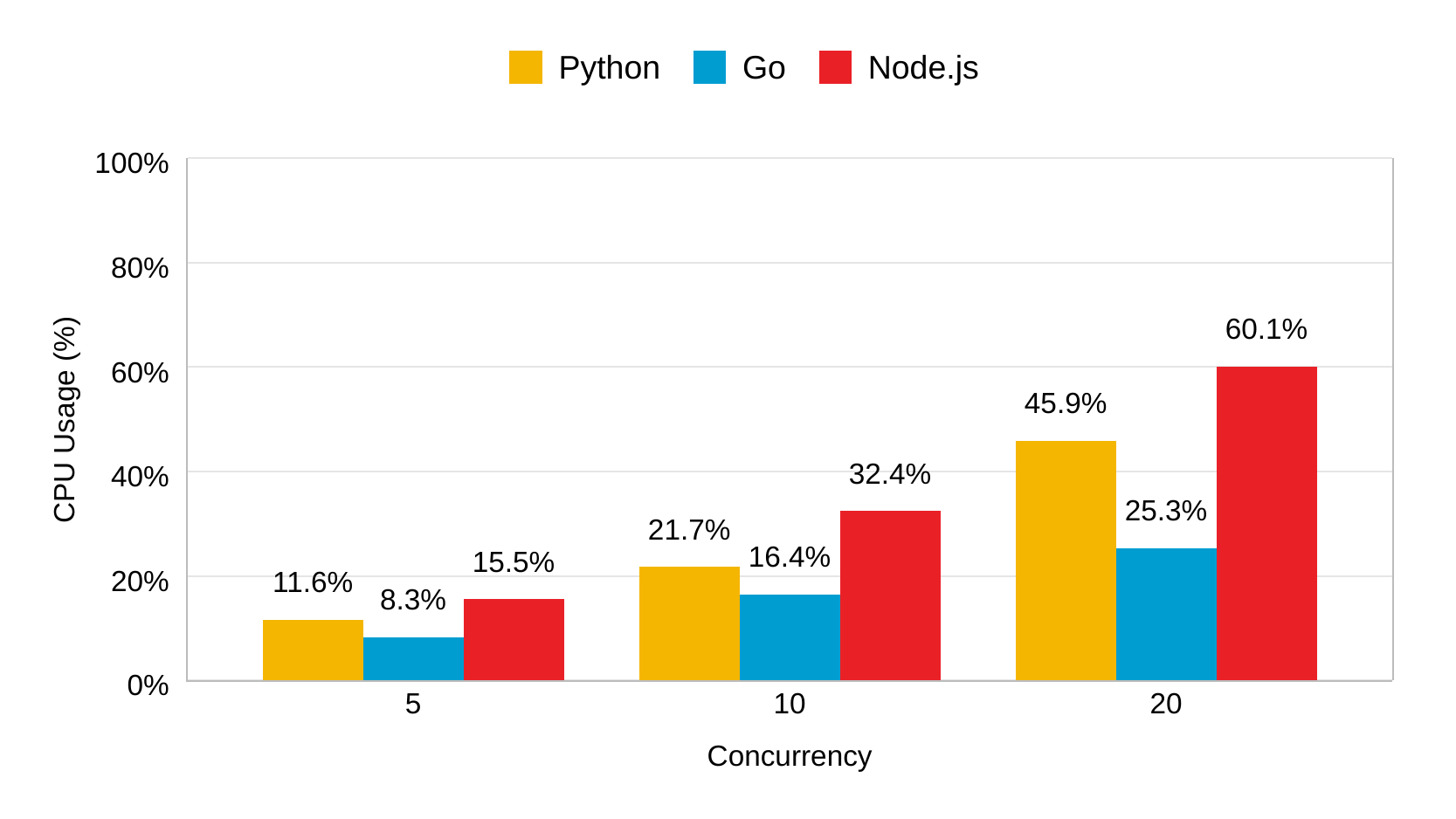}
		\caption{K3s CPU usage}
		\label{fig12}
	\end{minipage}
\end{figure}

\begin{figure}[!h]
	\centering
	\begin{minipage}{0.48\textwidth}
		\centering
		\includegraphics[width=\textwidth]{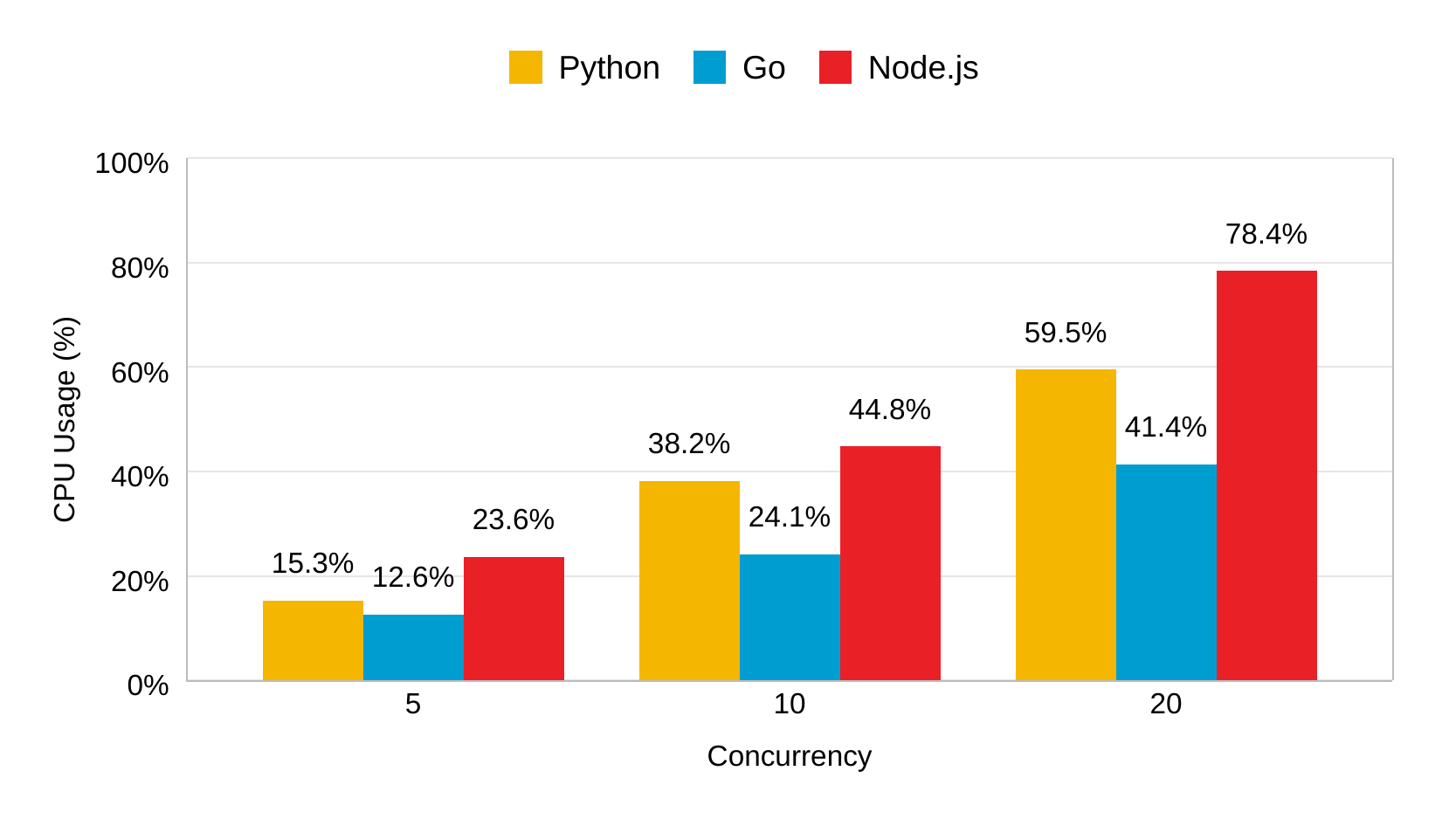}
		\caption{MicroK8s CPU usage}
		\label{fig13}
	\end{minipage}%
	\hfill
	\begin{minipage}{0.48\textwidth}
		\centering
		\includegraphics[width=\textwidth]{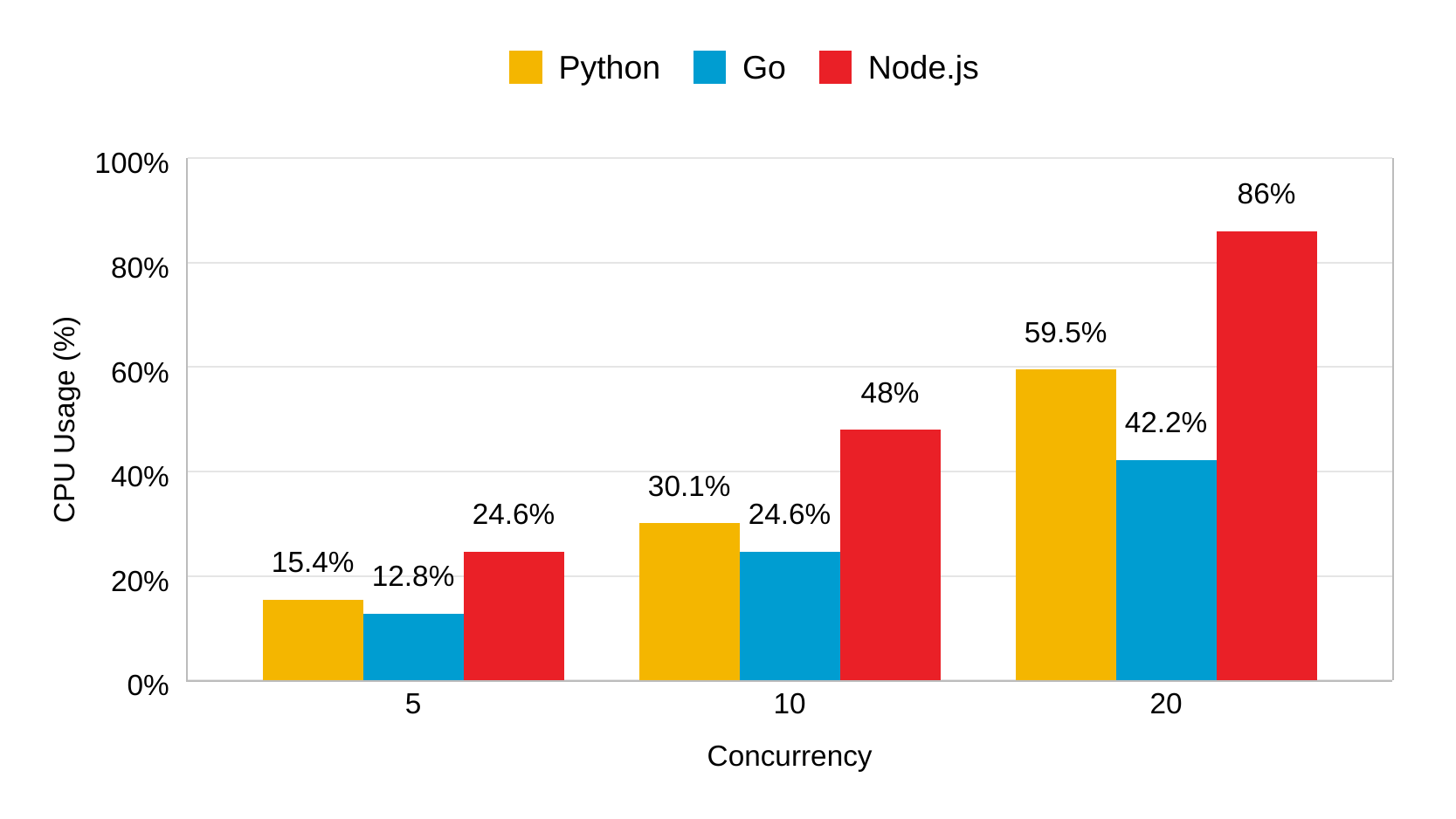}
		\caption{K0s CPU usage}
		\label{fig14}
	\end{minipage}
\end{figure}

In the previous part, we discussed how Go has the best performance among programming language runtimes, but what about its CPU usage? Figures \ref{fig11} to \ref{fig14} illustrate the CPU usage of programming language runtimes for K8s distributions.

Based on the data, all languages exhibit a progressive increase in CPU usage as the number of concurrent operations grows. The trend indicates that Node.js is more resource-intensive than Python and Go, with Python following in second place and Go ranked third. This result highlights Go's more efficient resource management and its ability to handle higher workloads in serverless environments without a proportionate increase in CPU load.
Furthermore, no specific language appears to be more sensitive to concurrency levels than others in terms of CPU usage. From the perspective of Kubernetes distributions, regardless of the programming language or concurrency level, Kubeadm, K3s, MicroK8s, and K0s rank first to fourth in terms of CPU usage, respectively. In other words, Kubeadm consumes the least amount of CPU compared to the other distributions when integrated with OpenFaaS to run our function invocations.

\subsection{Kubernetes Distribution}\label{subsec4-3}

\begin{figure}[!h]
	\centering
	\begin{minipage}{0.48\textwidth}
		\centering
		\includegraphics[width=\textwidth]{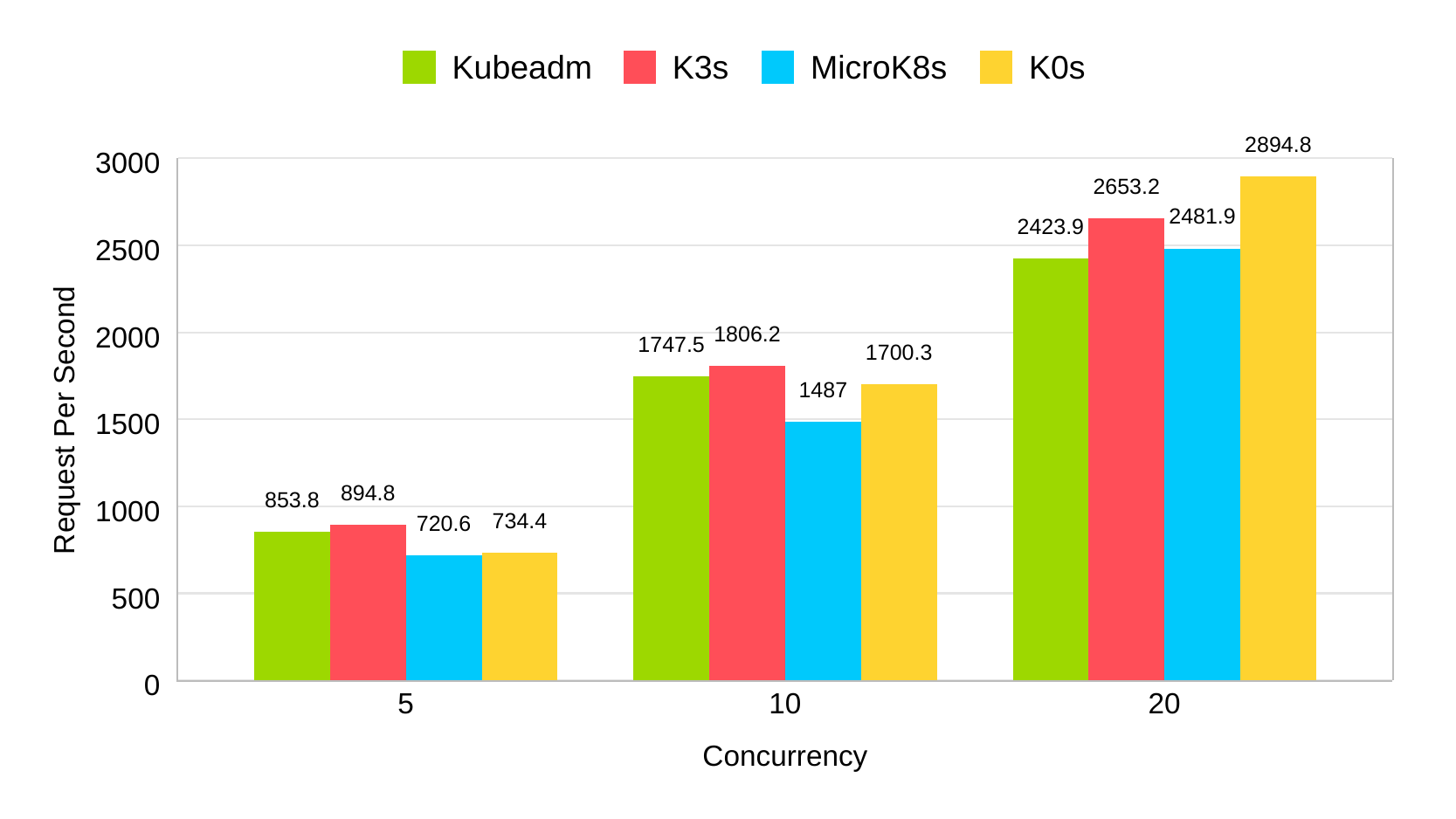}
		\caption{Request per second for Kubernetes distributions}
		\label{fig15}
	\end{minipage}%
	\hfill
	\begin{minipage}{0.48\textwidth}
		\centering
		\includegraphics[width=\textwidth]{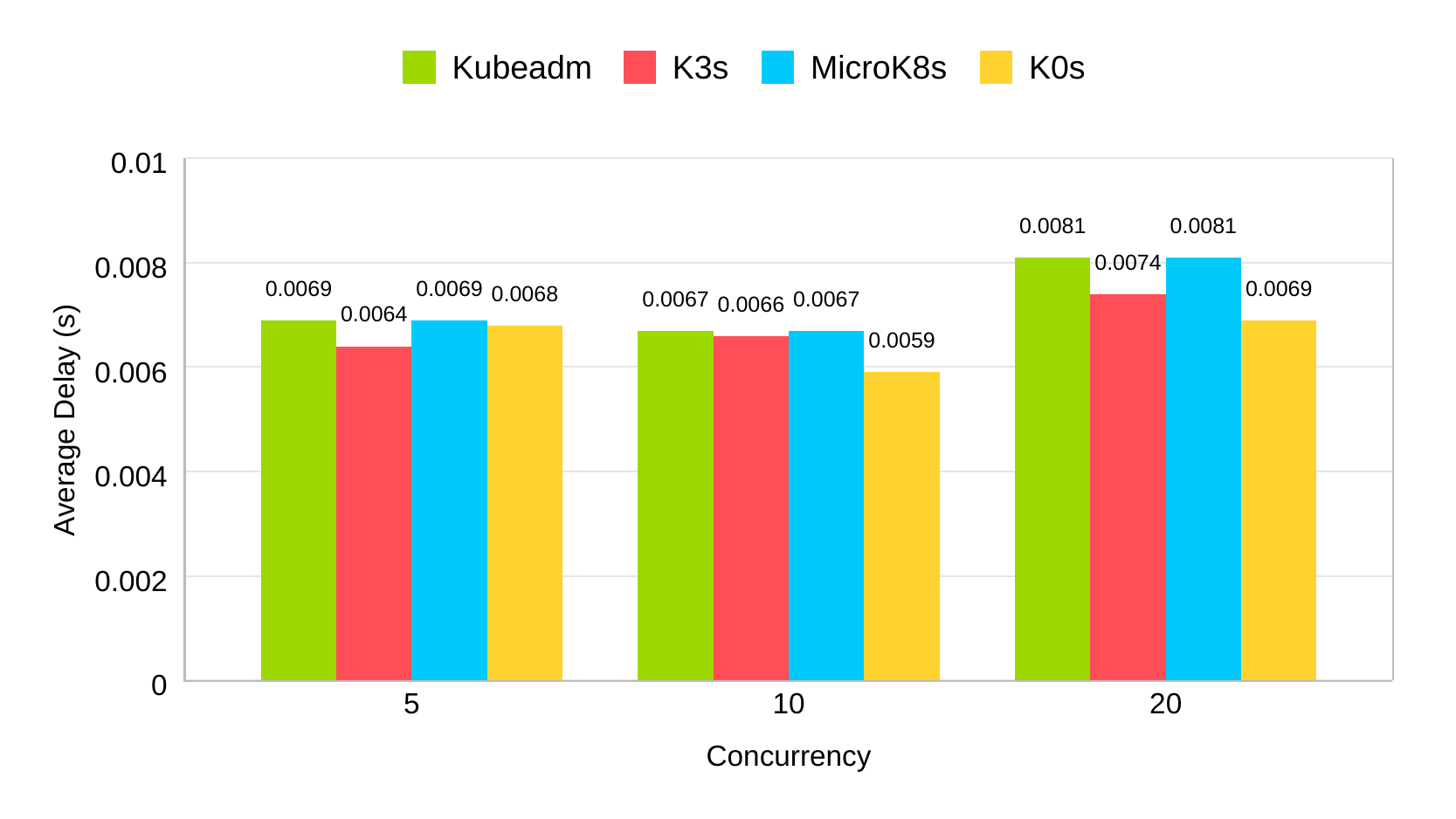}
		\caption{Average delay for Kubernetes distributions}
		\label{fig16}
	\end{minipage}
\end{figure}

\begin{center}
	\begin{figure}[!h]
		\centering
		\includegraphics[width=0.48 \textwidth]{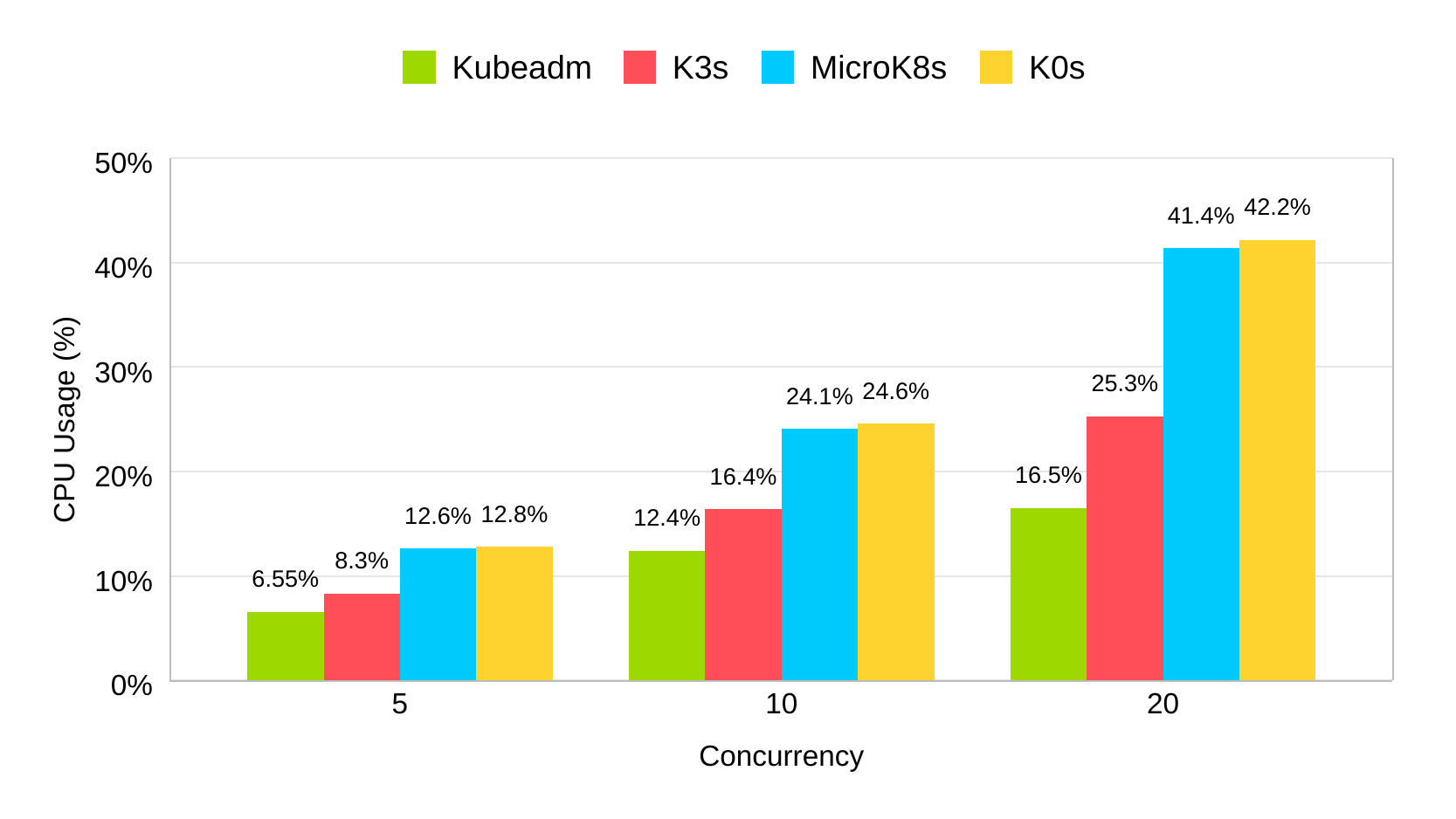}
		\caption{CPU usage for Kubernetes distributions }
		\label{fig17}
	\end{figure}
\end{center}

A comparison of RPS,  average delay, and CPU usage across K8s distributions including, Kubeadm, K3s, MicroK8s, and K0s reveals notable differences in performance and usage metrics with differences in concurrency levels. Figures \ref{fig15} to \ref{fig17} illustrate the difference in these performance measures between K8s distributions for Go programming language. Here, Go is considered to develop the functions deployed on OpenFaaS because it shows the best performance results in the experiments of Sections \ref{subsec4-1} and \ref{subsec4-2}. 

Based on RPS, both Kubeadm and K3s exhibit impressive throughput capabilities, showing strong scalability as concurrency increases. With consistently high RPS without significant performance drops, they handle growing demands effectively. In other words, both Kubeadm and K3s are capable of handling multiple requests simultaneously. On the other hand, while MicroK8s and K0s exhibit slightly lower performance at lower concurrency levels, they deliver strong RPS performance at higher concurrency levels and adapt well to increasing loads. 

on delay, at a concurrency level of 5, K3s provides the best performance, with K0s in second place. However, at concurrency levels of 10 and 15, K0s takes the lead, exhibiting the lowest average delay, while K3s ranks second. Surprisingly, Kubeadm and MicroK8s perform similarly, consistently tying for third place, regardless of the number of concurrent function invocations.

Based on CPU usage, Kubeadm consumes the least amount of CPU among all distributions. As concurrency rises, Kubeadm shows a moderate increase in CPU usage compared to other distributions, indicating efficient CPU management and ensuring consistent performance without straining resources. K3s ranks second, consuming 26\%, 32\%, and 53\% more CPU than Kubeadm at concurrency levels of 5, 10, and 20, respectively. These values indicate that K3s is more sensitive to concurrency increases than Kubeadm. MicroK8s and K0s behave almost similarly, ranking third and fourth, respectively. On average, these two distributions consume 94\%, 97\%, and 153\% more CPU than Kubeadm at concurrency levels of 5, 10, and 20, respectively, demonstrating that at higher concurrencies, they struggle to manage CPU resources efficiently.

Although all four K8s distributions maintain strong performance metrics, Kubeadm and K3s generally offer a more balanced approach between resource efficiency and high throughput. MicroK8s and K0s, on the other hand, may consume more resources, affecting overall efficiency in processor-intensive scenarios. Based on specific operational requirements and resource availability, a nuanced understanding of the performance of each distribution helps determine which distribution would be most suitable for the specific task at hand.

\subsection{Discussion}\label{subsec4-4}
As shown in previous sections, from a programming language perspective, Go outperforms other languages in terms of RPS and delay, regardless of the Kubernetes distribution or concurrency level. Performance trends observed in Python, Go, and Node.js on the OpenFaaS serverless platform can be attributed to several key factors inherent to these programming language runtimes and their runtime environments: \\
\textbf{\textit{Parallelism Handling}:} Go is known for its efficient concurrency model, utilizing lightweight threads called \textit{goroutines}. Due to this feature, Go can efficiently handle a large number of concurrent processes. OpenFaaS and other serverless environments like it benefit from the memory-efficient nature of goroutines. In Node.js, the I/O model is event-driven, non-blocking, and typically runs in a single-threaded environment. Especially on serverless platforms, Node.js can cause performance bottlenecks for CPU-intensive tasks, resulting in increased delays as concurrency increases. The \textit{Global interpreter lock} (GIL) limits the execution of Python bytecode to one thread at a time, which restricts its performance. Python supports concurrency via threads and asynchronous programming. Due to this characteristic, Python cannot perform optimally in multi-threaded scenarios, crucial for serverless platforms with high concurrency levels. \\	
\textbf{\textit{Optimization}:} 
	On serverless platforms where efficient execution is essential, Go's compilation to machine code optimizes system resource usage. During peak demand, this efficiency allows for higher throughput and quicker average delay. As interpreted languages, Node.js and Python typically exhibit slower execution speeds than compiled languages like Go. Due to this intrinsic characteristic and their concurrency handling, a serverless platform may have increased average delay under high concurrency. \\	
\textbf{\textit{Memory Management}:} 
	Despite varying loads in a serverless setting, Go offers performance stability due to the advanced garbage collection and statically typed system. With Go's advanced garbage collection, unused resources are promptly released, ensuring that memory is efficiently allocated and dealt with. This prevents memory leaks and reduces the likelihood of performance degradation, resulting in stable and consistent performance even in the face of varying loads in a serverless environment. In serverless workloads with heavy and fluctuating memory usage, Node.js and Python suffer from less efficient memory management and garbage collection.

\begin{figure}[b]
	\centering
	\includegraphics[width=1\textwidth]{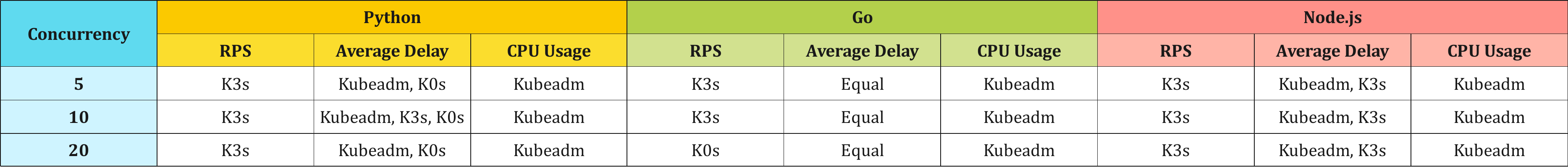}
	\captionsetup{justification=centering}
	\caption{Winners of Kubernetes distributions by programming language and performance metrics}\label{fig18}
\end{figure}

Figure \ref{fig18} compares the performance of different K8s distributions (K3s, Kubeadm, MicroK8s, and K0s) in terms of RPS, average delay, and CPU usage at different concurrency levels. According to these rankings, each distribution excels in specific performance metrics based on its architecture, which can guide selection based on application requirements. 
With regards to RPS, K3s consistently leads across all concurrency levels for Python and Node.js, demonstrating exceptional efficiency when handling large volumes of requests. K3s also maintains a strong position in Go, sharing the top spot with K0s at higher concurrency levels. 
Kubeadm excels at maintaining low  delay across all concurrency levels for Python, Go, and Node.js. This shows how Kubeadm is optimized for rapid response time, letting applications remain responsive even under heavy load. It also shares the top rank especially with K3s and K0s in some scenarios. Furthermore, Kubeadm ranks highest in CPU efficiency, ensuring low  consumption of processing resources across Python, Go, and Node.js. As a result, users benefit from high performance with minimal overhead due to Kubeadm's efficient use of computing resources.

\section{Conclusion and Future work}\label{sec5}

Various Kubernetes distributions, including Kubeadm, K3s, MicroK8s, and K0s, were investigated in this paper to compare the performance of OpenFaaS serverless platform when integrated with Kubernetes and the programming language runtimes when they are used to develop functions for OpenFaas. Using Python, Go, and Node.js programming language runtimes, we analyzed these distributions based on requests per second, average delay, and CPU usage. 
This integrated analysis of RPS, delay, and processor usage provides valuable insights into the unique strengths of each distribution and programming language, enabling informed decisions based on the specific application requirements for each environment.

Based on the experimental results, K3s and Kubeadm emerged as the leading platforms, with K3s excelling in providing high throughput and Kubeadm excelling in maintaining the lowest average delay while efficiently managing CPU resources.
Among the programming language runtimes, Go delivered high performance with minimal delay and efficient CPU usage, while Python provided a moderate balance between performance and resource consumption. However, Node.js exhibited higher CPU usage and longer average delays under high concurrency. 

As future work, we plan to combine machine learning models with serverless platforms and container orchestration tools to adapt to varying workloads and dynamically optimize resource allocation. For instance, adjusting the number of prewarmed containers based on the historical number of concurrent invocations could help reduce cold start delays. Additionally, utilizing reinforcement learning techniques could further minimize the occurrence of cold starts. These approaches are expected to lead to better resource utilization and improved overall performance. 
Another potential direction for future work could be to propose analytical models, such as queueing networks or Petri nets, to model the behavior and performance of OpenFaaS when integrated with Kubernetes. This approach would allow the results of the analytical modeling to be validated against the experimental findings provided in this study. Once validated, the proposed analytical models could be leveraged for extended analysis, including what-if scenarios.

\section*{Acknowledgement}
This research was financed by a research grant from the University of Mazandaran.

\bibliography{sn-bibliography}


\begin{thebibliography}{61}
\ifx \bisbn   \undefined \def \bisbn  #1{ISBN #1}\fi
\ifx \binits  \undefined \def \binits#1{#1}\fi
\ifx \bauthor  \undefined \def \bauthor#1{#1}\fi
\ifx \batitle  \undefined \def \batitle#1{#1}\fi
\ifx \bjtitle  \undefined \def \bjtitle#1{#1}\fi
\ifx \bvolume  \undefined \def \bvolume#1{\textbf{#1}}\fi
\ifx \byear  \undefined \def \byear#1{#1}\fi
\ifx \bissue  \undefined \def \bissue#1{#1}\fi
\ifx \bfpage  \undefined \def \bfpage#1{#1}\fi
\ifx \blpage  \undefined \def \blpage #1{#1}\fi
\ifx \burl  \undefined \def \burl#1{\textsf{#1}}\fi
\ifx \doiurl  \undefined \def \doiurl#1{\url{https://doi.org/#1}}\fi
\ifx \betal  \undefined \def \betal{\textit{et al.}}\fi
\ifx \binstitute  \undefined \def \binstitute#1{#1}\fi
\ifx \binstitutionaled  \undefined \def \binstitutionaled#1{#1}\fi
\ifx \bctitle  \undefined \def \bctitle#1{#1}\fi
\ifx \beditor  \undefined \def \beditor#1{#1}\fi
\ifx \bpublisher  \undefined \def \bpublisher#1{#1}\fi
\ifx \bbtitle  \undefined \def \bbtitle#1{#1}\fi
\ifx \bedition  \undefined \def \bedition#1{#1}\fi
\ifx \bseriesno  \undefined \def \bseriesno#1{#1}\fi
\ifx \blocation  \undefined \def \blocation#1{#1}\fi
\ifx \bsertitle  \undefined \def \bsertitle#1{#1}\fi
\ifx \bsnm \undefined \def \bsnm#1{#1}\fi
\ifx \bsuffix \undefined \def \bsuffix#1{#1}\fi
\ifx \bparticle \undefined \def \bparticle#1{#1}\fi
\ifx \barticle \undefined \def \barticle#1{#1}\fi
\bibcommenthead
\ifx \bconfdate \undefined \def \bconfdate #1{#1}\fi
\ifx \botherref \undefined \def \botherref #1{#1}\fi
\ifx \url \undefined \def \url#1{\textsf{#1}}\fi
\ifx \bchapter \undefined \def \bchapter#1{#1}\fi
\ifx \bbook \undefined \def \bbook#1{#1}\fi
\ifx \bcomment \undefined \def \bcomment#1{#1}\fi
\ifx \oauthor \undefined \def \oauthor#1{#1}\fi
\ifx \citeauthoryear \undefined \def \citeauthoryear#1{#1}\fi
\ifx \endbibitem  \undefined \def \endbibitem {}\fi
\ifx \bconflocation  \undefined \def \bconflocation#1{#1}\fi
\ifx \arxivurl  \undefined \def \arxivurl#1{\textsf{#1}}\fi
\csname PreBibitemsHook\endcsname

\bibitem[\protect\citeauthoryear{Armbrust et~al.}{2009}]{armbrust2009above}
\begin{botherref}
\oauthor{\bsnm{Armbrust}, \binits{M.}},
\oauthor{\bsnm{Fox}, \binits{A.}},
\oauthor{\bsnm{Griffith}, \binits{R.}},
\oauthor{\bsnm{Joseph}, \binits{A.D.}},
\oauthor{\bsnm{Katz}, \binits{R.H.}},
\oauthor{\bsnm{Konwinski}, \binits{A.}},
\oauthor{\bsnm{Lee}, \binits{G.}},
\oauthor{\bsnm{Patterson}, \binits{D.A.}},
\oauthor{\bsnm{Rabkin}, \binits{A.}},
\oauthor{\bsnm{Stoica}, \binits{I.}}, et al.:
Above the clouds: A berkeley view of cloud computing.
Technical report,
Technical Report UCB/EECS-2009-28, EECS Department, University of California
(2009)
\end{botherref}
\endbibitem

\bibitem[\protect\citeauthoryear{Aqasizade
  et~al.}{2024a}]{aqasizade2024kubernetes}
\begin{botherref}
\oauthor{\bsnm{Aqasizade}, \binits{H.}},
\oauthor{\bsnm{Ataie}, \binits{E.}},
\oauthor{\bsnm{Bastam}, \binits{M.}}:
Kubernetes in action: Exploring the performance of kubernetes distributions in
  the cloud.
arXiv preprint arXiv:2403.01429
(2024)
\end{botherref}
\endbibitem

\bibitem[\protect\citeauthoryear{Aqasizade
  et~al.}{2024b}]{aqasizade2024experimental}
\begin{botherref}
\oauthor{\bsnm{Aqasizade}, \binits{H.}},
\oauthor{\bsnm{Ataie}, \binits{E.}},
\oauthor{\bsnm{Bastam}, \binits{M.}}:
Experimental assessment of containers running on top of virtual machines.
arXiv preprint arXiv:2401.07539
(2024)
\end{botherref}
\endbibitem

\bibitem[\protect\citeauthoryear{Risco et~al.}{2021}]{risco2021serverless}
\begin{barticle}
\bauthor{\bsnm{Risco}, \binits{S.}},
\bauthor{\bsnm{Molt{\'o}}, \binits{G.}},
\bauthor{\bsnm{Naranjo}, \binits{D.M.}},
\bauthor{\bsnm{Blanquer}, \binits{I.}}:
\batitle{Serverless workflows for containerised applications in the cloud
  continuum}.
\bjtitle{Journal of Grid Computing}
\bvolume{19},
\bfpage{1}--\blpage{18}
(\byear{2021})
\end{barticle}
\endbibitem

\bibitem[\protect\citeauthoryear{Shojaee~rad et~al.}{2024}]{shojaee2024memory}
\begin{botherref}
\oauthor{\bsnm{Shojaee~rad}, \binits{Z.}},
\oauthor{\bsnm{Ghobaei-Arani}, \binits{M.}},
\oauthor{\bsnm{Ahsan}, \binits{R.}}:
Memory orchestration mechanisms in serverless computing: a taxonomy, review and
  future directions.
Cluster Computing,
1--27
(2024)
\end{botherref}
\endbibitem

\bibitem[\protect\citeauthoryear{Scheuner and
  Leitner}{2020}]{scheuner2020function}
\begin{barticle}
\bauthor{\bsnm{Scheuner}, \binits{J.}},
\bauthor{\bsnm{Leitner}, \binits{P.}}:
\batitle{Function-as-a-service performance evaluation: A multivocal literature
  review}.
\bjtitle{Journal of Systems and Software}
\bvolume{170},
\bfpage{110708}
(\byear{2020})
\end{barticle}
\endbibitem

\bibitem[\protect\citeauthoryear{Yussupov et~al.}{2021}]{yussupov2021faasten}
\begin{barticle}
\bauthor{\bsnm{Yussupov}, \binits{V.}},
\bauthor{\bsnm{Soldani}, \binits{J.}},
\bauthor{\bsnm{Breitenb{\"u}cher}, \binits{U.}},
\bauthor{\bsnm{Brogi}, \binits{A.}},
\bauthor{\bsnm{Leymann}, \binits{F.}}:
\batitle{Faasten your decisions: A classification framework and technology
  review of function-as-a-service platforms}.
\bjtitle{Journal of Systems and Software}
\bvolume{175},
\bfpage{110906}
(\byear{2021})
\end{barticle}
\endbibitem

\bibitem[\protect\citeauthoryear{Bouizem et~al.}{2023}]{bouizem2023integrating}
\begin{barticle}
\bauthor{\bsnm{Bouizem}, \binits{Y.}},
\bauthor{\bsnm{Dib}, \binits{D.}},
\bauthor{\bsnm{Parlavantzas}, \binits{N.}},
\bauthor{\bsnm{Morin}, \binits{C.}}:
\batitle{Integrating request replication into faas platforms: an experimental
  evaluation}.
\bjtitle{Journal of Cloud Computing}
\bvolume{12}(\bissue{1}),
\bfpage{94}
(\byear{2023})
\end{barticle}
\endbibitem

\bibitem[\protect\citeauthoryear{Ghorbian et~al.}{2024}]{ghorbian2024survey}
\begin{botherref}
\oauthor{\bsnm{Ghorbian}, \binits{M.}},
\oauthor{\bsnm{Ghobaei-Arani}, \binits{M.}},
\oauthor{\bsnm{Esmaeili}, \binits{L.}}:
A survey on the scheduling mechanisms in serverless computing: a taxonomy,
  challenges, and trends.
Cluster Computing,
1--40
(2024)
\end{botherref}
\endbibitem

\bibitem[\protect\citeauthoryear{Kumari and Sahoo}{2024}]{kumari2024acpm}
\begin{barticle}
\bauthor{\bsnm{Kumari}, \binits{A.}},
\bauthor{\bsnm{Sahoo}, \binits{B.}}:
\batitle{{ACPM}: adaptive container provisioning model to mitigate serverless
  cold-start}.
\bjtitle{Cluster Computing}
\bvolume{27}(\bissue{2}),
\bfpage{1333}--\blpage{1360}
(\byear{2024})
\end{barticle}
\endbibitem

\bibitem[\protect\citeauthoryear{}{}]{lambda}
\begin{botherref}
Amazon Lambda.
\url{https://aws.amazon.com/lambda}.
Accessed: September 2024
\end{botherref}
\endbibitem

\bibitem[\protect\citeauthoryear{}{}]{azure}
\begin{botherref}
Azure Serverless.
\url{https://azure.microsoft.com/en-us/free/serverless}.
Accessed: September 2024
\end{botherref}
\endbibitem

\bibitem[\protect\citeauthoryear{}{}]{GoogleCloudFunctions}
\begin{botherref}
Google Cloud Functions.
\url{https://cloud.google.com/functions/}.
Accessed: September 2024
\end{botherref}
\endbibitem

\bibitem[\protect\citeauthoryear{}{}]{IBMCloudFunctions}
\begin{botherref}
IBM Cloud Functions.
\url{https://https://cloud.ibm.com/functions/}.
Accessed: September 2024
\end{botherref}
\endbibitem

\bibitem[\protect\citeauthoryear{}{}]{OpenFaaS}
\begin{botherref}
OpenFaaS.
\url{https://www.openfaas.com}.
Accessed: September 2024
\end{botherref}
\endbibitem

\bibitem[\protect\citeauthoryear{}{}]{openwhisk}
\begin{botherref}
Apache OpenWhisk.
\url{https://openwhisk.apache.org}.
Accessed: September 2024
\end{botherref}
\endbibitem

\bibitem[\protect\citeauthoryear{}{}]{Kubeless}
\begin{botherref}
Kubeless.
\url{https://github.com/vmware-archive/kubeless}.
Accessed: September 2024
\end{botherref}
\endbibitem

\bibitem[\protect\citeauthoryear{}{}]{nuclio}
\begin{botherref}
Nuclio.
\url{https://nuclio.io}.
Accessed: September 2024
\end{botherref}
\endbibitem

\bibitem[\protect\citeauthoryear{}{}]{fission}
\begin{botherref}
Fission.
\url{https://fission.io}.
Accessed: September 2024
\end{botherref}
\endbibitem

\bibitem[\protect\citeauthoryear{}{}]{knative}
\begin{botherref}
Knative.
\url{https://knative.dev}.
Accessed: September 2024
\end{botherref}
\endbibitem

\bibitem[\protect\citeauthoryear{Liu et~al.}{2024}]{liu2024balanced}
\begin{botherref}
\oauthor{\bsnm{Liu}, \binits{J.}},
\oauthor{\bsnm{Ding}, \binits{Y.}},
\oauthor{\bsnm{Liu}, \binits{Y.}}:
A balanced leader election algorithm based on replica distribution in
  kubernetes cluster.
Cluster Computing,
1--10
(2024)
\end{botherref}
\endbibitem

\bibitem[\protect\citeauthoryear{}{}]{Kubeadm}
\begin{botherref}
Kubeadm.
\url{https://kubernetes.io/docs/reference/setup-tools/kubeadm/}.
Accessed: September 2024
\end{botherref}
\endbibitem

\bibitem[\protect\citeauthoryear{}{}]{k3s}
\begin{botherref}
K3s lightweight kubernetes.
\url{https://www.k3s.io}.
Accessed: September 2024
\end{botherref}
\endbibitem

\bibitem[\protect\citeauthoryear{}{}]{microk8s}
\begin{botherref}
microk8s.
\url{https://microk8s.io/}.
Accessed: September 2024
\end{botherref}
\endbibitem

\bibitem[\protect\citeauthoryear{}{}]{k0s}
\begin{botherref}
k0s project.
\url{https://k0sproject.io}.
Accessed: September 2024
\end{botherref}
\endbibitem

\bibitem[\protect\citeauthoryear{Chahoud et~al.}{2023}]{chahoud2023demand}
\begin{botherref}
\oauthor{\bsnm{Chahoud}, \binits{M.}},
\oauthor{\bsnm{Sami}, \binits{H.}},
\oauthor{\bsnm{Mourad}, \binits{A.}},
\oauthor{\bsnm{Otoum}, \binits{S.}},
\oauthor{\bsnm{Otrok}, \binits{H.}},
\oauthor{\bsnm{Bentahar}, \binits{J.}},
\oauthor{\bsnm{Guizani}, \binits{M.}}:
On-demand-fl: A dynamic and efficient multi-criteria federated learning client
  deployment scheme.
IEEE Internet of Things Journal
(2023)
\end{botherref}
\endbibitem

\bibitem[\protect\citeauthoryear{Basak and Srirama}{2023}]{basakfog}
\begin{botherref}
\oauthor{\bsnm{Basak}, \binits{S.}},
\oauthor{\bsnm{Srirama}, \binits{S.N.}}:
Fog computing out of the box: Dynamic deployment of fog service containers with
  tosca.
International Journal of Network Management,
2246
(2023)
\end{botherref}
\endbibitem

\bibitem[\protect\citeauthoryear{Koziolek and
  Eskandani}{2023}]{koziolek2023lightweight}
\begin{bchapter}
\bauthor{\bsnm{Koziolek}, \binits{H.}},
\bauthor{\bsnm{Eskandani}, \binits{N.}}:
\bctitle{Lightweight kubernetes distributions: A performance comparison of
  microk8s, k3s, k0s, and microshift}.
In: \bbtitle{Proceedings of the 2023 ACM/SPEC International Conference on
  Performance Engineering},
pp. \bfpage{17}--\blpage{29}
(\byear{2023})
\end{bchapter}
\endbibitem

\bibitem[\protect\citeauthoryear{Heckmann and
  Ravindran}{2023}]{heckmann2023evaluating}
\begin{bchapter}
\bauthor{\bsnm{Heckmann}, \binits{O.}},
\bauthor{\bsnm{Ravindran}, \binits{A.}}:
\bctitle{Evaluating kubernetes at the edge for fault tolerant multi-camera
  computer vision applications}.
In: \bbtitle{2023 IEEE/ACM 23rd International Symposium on Cluster, Cloud and
  Internet Computing Workshops (CCGridW)},
pp. \bfpage{269}--\blpage{271}
(\byear{2023})
\end{bchapter}
\endbibitem

\bibitem[\protect\citeauthoryear{Kim et~al.}{2023}]{kim2023device}
\begin{barticle}
\bauthor{\bsnm{Kim}, \binits{B.}},
\bauthor{\bsnm{Calin}, \binits{D.}},
\bauthor{\bsnm{Tenny}, \binits{N.}},
\bauthor{\bsnm{Shariat}, \binits{M.}},
\bauthor{\bsnm{Fan}, \binits{M.}}:
\batitle{Device centric distributed compute, orchestration and networking}.
\bjtitle{IEEE Wireless Communications}
\bvolume{30}(\bissue{4}),
\bfpage{6}--\blpage{8}
(\byear{2023})
\end{barticle}
\endbibitem

\bibitem[\protect\citeauthoryear{Duplyakin et~al.}{2019}]{duplyakin2019design}
\begin{bchapter}
\bauthor{\bsnm{Duplyakin}, \binits{D.}},
\bauthor{\bsnm{Ricci}, \binits{R.}},
\bauthor{\bsnm{Maricq}, \binits{A.}},
\bauthor{\bsnm{Wong}, \binits{G.}},
\bauthor{\bsnm{Duerig}, \binits{J.}},
\bauthor{\bsnm{Eide}, \binits{E.}},
\bauthor{\bsnm{Stoller}, \binits{L.}},
\bauthor{\bsnm{Hibler}, \binits{M.}},
\bauthor{\bsnm{Johnson}, \binits{D.}},
\bauthor{\bsnm{Webb}, \binits{K.}}, \betal:
\bctitle{The design and operation of $\{$CloudLab$\}$}.
In: \bbtitle{2019 USENIX Annual Technical Conference (USENIX ATC 19)},
pp. \bfpage{1}--\blpage{14}
(\byear{2019})
\end{bchapter}
\endbibitem

\bibitem[\protect\citeauthoryear{Ramanathan
  et~al.}{2024}]{ramanathan2024enabling}
\begin{barticle}
\bauthor{\bsnm{Ramanathan}, \binits{S.}},
\bauthor{\bsnm{Bhattacharyya}, \binits{A.}},
\bauthor{\bsnm{Kondepu}, \binits{K.}},
\bauthor{\bsnm{Fumagalli}, \binits{A.}}:
\batitle{Enabling containerized central unit live migration in 5g radio access
  network: An experimental study}.
\bjtitle{Journal of Network and Computer Applications}
\bvolume{221},
\bfpage{103767}
(\byear{2024})
\end{barticle}
\endbibitem

\bibitem[\protect\citeauthoryear{Ganguly et~al.}{2023}]{ganguly2023multi}
\begin{botherref}
\oauthor{\bsnm{Ganguly}, \binits{B.}},
\oauthor{\bsnm{Hosseinalipour}, \binits{S.}},
\oauthor{\bsnm{Kim}, \binits{K.T.}},
\oauthor{\bsnm{Brinton}, \binits{C.G.}},
\oauthor{\bsnm{Aggarwal}, \binits{V.}},
\oauthor{\bsnm{Love}, \binits{D.J.}},
\oauthor{\bsnm{Chiang}, \binits{M.}}:
Multi-edge server-assisted dynamic federated learning with an optimized
  floating aggregation point.
IEEE/ACM Transactions on Networking
(2023)
\end{botherref}
\endbibitem

\bibitem[\protect\citeauthoryear{Qi et~al.}{2023}]{qi2023middlenet}
\begin{botherref}
\oauthor{\bsnm{Qi}, \binits{S.}},
\oauthor{\bsnm{Zeng}, \binits{Z.}},
\oauthor{\bsnm{Monis}, \binits{L.}},
\oauthor{\bsnm{Ramakrishnan}, \binits{K.}}:
Middlenet: A unified, high-performance nfv and middlebox framework with ebpf
  and dpdk.
IEEE Transactions on Network and Service Management
(2023)
\end{botherref}
\endbibitem

\bibitem[\protect\citeauthoryear{}{}]{hey}
\begin{botherref}
Hey.
\url{https://github.com/rakyll/hey}.
Accessed: September 2024
\end{botherref}
\endbibitem

\bibitem[\protect\citeauthoryear{}{}]{k6}
\begin{botherref}
Grafana k6.
\url{https://k6.io}.
Accessed: September 2024
\end{botherref}
\endbibitem

\bibitem[\protect\citeauthoryear{Kjorveziroski and
  Filiposka}{2022}]{kjorveziroski2022kubernetes}
\begin{barticle}
\bauthor{\bsnm{Kjorveziroski}, \binits{V.}},
\bauthor{\bsnm{Filiposka}, \binits{S.}}:
\batitle{Kubernetes distributions for the edge: serverless performance
  evaluation}.
\bjtitle{The Journal of Supercomputing}
\bvolume{78}(\bissue{11}),
\bfpage{13728}--\blpage{13755}
(\byear{2022})
\end{barticle}
\endbibitem

\bibitem[\protect\citeauthoryear{Mohanty et~al.}{2018}]{mohanty2018evaluation}
\begin{bchapter}
\bauthor{\bsnm{Mohanty}, \binits{S.K.}},
\bauthor{\bsnm{Premsankar}, \binits{G.}},
\bauthor{\bsnm{Di~Francesco}, \binits{M.}}:
\bctitle{An evaluation of open source serverless computing frameworks}.
In: \bbtitle{IEEE International Conference on Cloud Computing Technology and
  Science},
pp. \bfpage{115}--\blpage{120}
(\byear{2018})
\end{bchapter}
\endbibitem

\bibitem[\protect\citeauthoryear{Palade et~al.}{2019}]{8817155}
\begin{bchapter}
\bauthor{\bsnm{Palade}, \binits{A.}},
\bauthor{\bsnm{Kazmi}, \binits{A.}},
\bauthor{\bsnm{Clarke}, \binits{S.}}:
\bctitle{An evaluation of open source serverless computing frameworks support
  at the edge}.
In: \bbtitle{2019 IEEE World Congress on Services (SERVICES)},
vol. \bseriesno{2642-939X},
pp. \bfpage{206}--\blpage{211}
(\byear{2019})
\end{bchapter}
\endbibitem

\bibitem[\protect\citeauthoryear{Balla et~al.}{2020}]{9163456}
\begin{bchapter}
\bauthor{\bsnm{Balla}, \binits{D.}},
\bauthor{\bsnm{Maliosz}, \binits{M.}},
\bauthor{\bsnm{Simon}, \binits{C.}}:
\bctitle{Open source faas performance aspects}.
In: \bbtitle{2020 43rd International Conference on Telecommunications and
  Signal Processing (TSP)},
pp. \bfpage{358}--\blpage{364}
(\byear{2020})
\end{bchapter}
\endbibitem

\bibitem[\protect\citeauthoryear{Li}{2021}]{Li_2021}
\begin{bchapter}
\bauthor{\bsnm{Li}, \binits{J.}}:
\bctitle{Analyzing open-source serverless platforms: Characteristics and
  performance (s)}.
In: \bbtitle{International Conferences on Software Engineering and Knowledge
  Engineering}.
\bsertitle{SEKE2021}
(\byear{2021})
\end{bchapter}
\endbibitem

\bibitem[\protect\citeauthoryear{Decker et~al.}{2022}]{a15070234}
\begin{botherref}
\oauthor{\bsnm{Decker}, \binits{J.}},
\oauthor{\bsnm{Kasprzak}, \binits{P.}},
\oauthor{\bsnm{Kunkel}, \binits{J.M.}}:
Performance evaluation of open-source serverless platforms for kubernetes.
Algorithms
\textbf{15}(7)
(2022)
\end{botherref}
\endbibitem

\bibitem[\protect\citeauthoryear{Lee et~al.}{2018}]{8457830}
\begin{bchapter}
\bauthor{\bsnm{Lee}, \binits{H.}},
\bauthor{\bsnm{Satyam}, \binits{K.}},
\bauthor{\bsnm{Fox}, \binits{G.}}:
\bctitle{Evaluation of production serverless computing environments}.
In: \bbtitle{2018 IEEE 11th International Conference on Cloud Computing
  (CLOUD)},
pp. \bfpage{442}--\blpage{450}
(\byear{2018})
\end{bchapter}
\endbibitem

\bibitem[\protect\citeauthoryear{Wen
  et~al.}{2023}]{https://doi.org/10.1002/smr.2394}
\begin{barticle}
\bauthor{\bsnm{Wen}, \binits{J.}},
\bauthor{\bsnm{Liu}, \binits{Y.}},
\bauthor{\bsnm{Chen}, \binits{Z.}},
\bauthor{\bsnm{Chen}, \binits{J.}},
\bauthor{\bsnm{Ma}, \binits{Y.}}:
\batitle{Characterizing commodity serverless computing platforms}.
\bjtitle{Journal of Software: Evolution and Process}
\bvolume{35}(\bissue{10}),
\bfpage{2394}
(\byear{2023})
\end{barticle}
\endbibitem

\bibitem[\protect\citeauthoryear{Yu et~al.}{2020}]{yu2020characterizing}
\begin{bchapter}
\bauthor{\bsnm{Yu}, \binits{T.}},
\bauthor{\bsnm{Liu}, \binits{Q.}},
\bauthor{\bsnm{Du}, \binits{D.}},
\bauthor{\bsnm{Xia}, \binits{Y.}},
\bauthor{\bsnm{Zang}, \binits{B.}},
\bauthor{\bsnm{Lu}, \binits{Z.}},
\bauthor{\bsnm{Yang}, \binits{P.}},
\bauthor{\bsnm{Qin}, \binits{C.}},
\bauthor{\bsnm{Chen}, \binits{H.}}:
\bctitle{Characterizing serverless platforms with serverlessbench}.
In: \bbtitle{Proceedings of the 11th ACM Symposium on Cloud Computing},
pp. \bfpage{30}--\blpage{44}
(\byear{2020})
\end{bchapter}
\endbibitem

\bibitem[\protect\citeauthoryear{Wang et~al.}{2018}]{wang2018peeking}
\begin{bchapter}
\bauthor{\bsnm{Wang}, \binits{L.}},
\bauthor{\bsnm{Li}, \binits{M.}},
\bauthor{\bsnm{Zhang}, \binits{Y.}},
\bauthor{\bsnm{Ristenpart}, \binits{T.}},
\bauthor{\bsnm{Swift}, \binits{M.}}:
\bctitle{Peeking behind the curtains of serverless platforms}.
In: \bbtitle{2018 USENIX Annual Technical Conference (USENIX ATC 18)},
pp. \bfpage{133}--\blpage{146}
(\byear{2018})
\end{bchapter}
\endbibitem

\bibitem[\protect\citeauthoryear{Figiela et~al.}{2018}]{figiela2018performance}
\begin{barticle}
\bauthor{\bsnm{Figiela}, \binits{K.}},
\bauthor{\bsnm{Gajek}, \binits{A.}},
\bauthor{\bsnm{Zima}, \binits{A.}},
\bauthor{\bsnm{Obrok}, \binits{B.}},
\bauthor{\bsnm{Malawski}, \binits{M.}}:
\batitle{Performance evaluation of heterogeneous cloud functions}.
\bjtitle{Concurrency and Computation: Practice and Experience}
\bvolume{30}(\bissue{23}),
\bfpage{4792}
(\byear{2018})
\end{barticle}
\endbibitem

\bibitem[\protect\citeauthoryear{Javed et~al.}{2022}]{javed2022serverless}
\begin{bchapter}
\bauthor{\bsnm{Javed}, \binits{H.}},
\bauthor{\bsnm{Toosi}, \binits{A.N.}},
\bauthor{\bsnm{Aslanpour}, \binits{M.S.}}:
\bctitle{Serverless platforms on the edge: a performance analysis}.
In: \bbtitle{New Frontiers in Cloud Computing and Internet of Things},
pp. \bfpage{165}--\blpage{184}
(\byear{2022})
\end{bchapter}
\endbibitem

\bibitem[\protect\citeauthoryear{Ataie et~al.}{2025}]{ataie2024}
\begin{barticle}
\bauthor{\bsnm{Ataie}, \binits{E.}},
\bauthor{\bsnm{Pooshani}, \binits{M.}},
\bauthor{\bsnm{Aqasizade}, \binits{H.}}:
\batitle{An empirical study on impact of programming languages on performance
  of open-source serverless platforms}.
\bjtitle{International Journal of Engineering}
\bvolume{38}(\bissue{2}),
\bfpage{424}--\blpage{435}
(\byear{2025})
\end{barticle}
\endbibitem

\bibitem[\protect\citeauthoryear{Goethals et~al.}{2019}]{goethals2019fledge}
\begin{bchapter}
\bauthor{\bsnm{Goethals}, \binits{T.}},
\bauthor{\bsnm{De~Turck}, \binits{F.}},
\bauthor{\bsnm{Volckaert}, \binits{B.}}:
\bctitle{Fledge: Kubernetes compatible container orchestration on low-resource
  edge devices}.
In: \bbtitle{International Conference on Internet of Vehicles},
pp. \bfpage{174}--\blpage{189}
(\byear{2019})
\end{bchapter}
\endbibitem

\bibitem[\protect\citeauthoryear{B{\"o}hm and Wirtz}{2021}]{bohm2021profiling}
\begin{bchapter}
\bauthor{\bsnm{B{\"o}hm}, \binits{S.}},
\bauthor{\bsnm{Wirtz}, \binits{G.}}:
\bctitle{Profiling lightweight container platforms: Microk8s and k3s in
  comparison to kubernetes.}
In: \bbtitle{ZEUS},
pp. \bfpage{65}--\blpage{73}
(\byear{2021})
\end{bchapter}
\endbibitem

\bibitem[\protect\citeauthoryear{Koukis et~al.}{2024}]{koukis2024performance}
\begin{botherref}
\oauthor{\bsnm{Koukis}, \binits{G.}},
\oauthor{\bsnm{Skaperas}, \binits{S.}},
\oauthor{\bsnm{Kapetanidou}, \binits{I.A.}},
\oauthor{\bsnm{Mamatas}, \binits{L.}},
\oauthor{\bsnm{Tsaoussidis}, \binits{V.}}:
Performance evaluation of kubernetes networking approaches across constraint
  edge environments.
arXiv preprint arXiv:2401.07674
(2024)
\end{botherref}
\endbibitem

\bibitem[\protect\citeauthoryear{Tzenetopoulos
  et~al.}{2021}]{tzenetopoulos2021faas}
\begin{bchapter}
\bauthor{\bsnm{Tzenetopoulos}, \binits{A.}},
\bauthor{\bsnm{Apostolakis}, \binits{E.}},
\bauthor{\bsnm{Tzomaka}, \binits{A.}},
\bauthor{\bsnm{Papakostopoulos}, \binits{C.}},
\bauthor{\bsnm{Stavrakakis}, \binits{K.}},
\bauthor{\bsnm{Katsaragakis}, \binits{M.}},
\bauthor{\bsnm{Oroutzoglou}, \binits{I.}},
\bauthor{\bsnm{Masouros}, \binits{D.}},
\bauthor{\bsnm{Xydis}, \binits{S.}},
\bauthor{\bsnm{Soudris}, \binits{D.}}:
\bctitle{Faas and curious: Performance implications of serverless functions on
  edge computing platforms}.
In: \bbtitle{High Performance Computing: ISC High Performance Digital 2021
  International Workshops, Frankfurt Am Main, Germany, June 24--July 2, 2021,
  Revised Selected Papers 36},
pp. \bfpage{428}--\blpage{438}
(\byear{2021})
\end{bchapter}
\endbibitem

\bibitem[\protect\citeauthoryear{Ustiugov et~al.}{2021}]{ustiugov2021analyzing}
\begin{bchapter}
\bauthor{\bsnm{Ustiugov}, \binits{D.}},
\bauthor{\bsnm{Amariucai}, \binits{T.}},
\bauthor{\bsnm{Grot}, \binits{B.}}:
\bctitle{Analyzing tail latency in serverless clouds with stellar}.
In: \bbtitle{2021 IEEE International Symposium on Workload Characterization
  (IISWC)},
pp. \bfpage{51}--\blpage{62}
(\byear{2021})
\end{bchapter}
\endbibitem

\bibitem[\protect\citeauthoryear{Trieu et~al.}{2022}]{trieu2022performance}
\begin{bchapter}
\bauthor{\bsnm{Trieu}, \binits{Q.L.}},
\bauthor{\bsnm{Javadi}, \binits{B.}},
\bauthor{\bsnm{Basilakis}, \binits{J.}},
\bauthor{\bsnm{Toosi}, \binits{A.N.}}:
\bctitle{Performance evaluation of serverless edge computing for machine
  learning applications}.
In: \bbtitle{2022 IEEE/ACM 15th International Conference on Utility and Cloud
  Computing (UCC)},
pp. \bfpage{139}--\blpage{144}
(\byear{2022})
\end{bchapter}
\endbibitem

\bibitem[\protect\citeauthoryear{}{}]{ubuntu}
\begin{botherref}
Enterprise Open source and Linux | Ubuntu.
\url{https://ubuntu.com}.
Accessed: September 2024
\end{botherref}
\endbibitem

\bibitem[\protect\citeauthoryear{}{}]{helm}
\begin{botherref}
Helm package manager.
\url{https://helm.sh}.
Accessed: September 2024
\end{botherref}
\endbibitem

\bibitem[\protect\citeauthoryear{}{}]{calico}
\begin{botherref}
Project Calico.
\url{https://www.tigera.io/project-calico}.
Accessed: September 2024
\end{botherref}
\endbibitem

\bibitem[\protect\citeauthoryear{}{}]{prometheus}
\begin{botherref}
Prometheus.
\url{https://prometheus.io}.
Accessed: September 2024
\end{botherref}
\endbibitem

\bibitem[\protect\citeauthoryear{}{}]{grafana}
\begin{botherref}
Grafana.
\url{https://grafana.com}.
Accessed: September 2024
\end{botherref}
\endbibitem

\bibitem[\protect\citeauthoryear{}{}]{pk8s}
\begin{botherref}
Prometheus k8s.
\url{https://github.com/aqasiz/Prometheus-k8s}.
Accessed: September 2024
\end{botherref}
\endbibitem

\end{thebibliography}

\end{document}